\documentclass[
  reprint,               
  superscriptaddress,    
  amsmath,amssymb,       
  aps,                   
  prb,            
  longbibliography       
]{revtex4-2}

\usepackage[english]{babel}
\usepackage{graphicx}   
\usepackage[colorlinks=true, allcolors=magenta]{hyperref}
\usepackage{braket}      
\usepackage{xcolor}      
\usepackage{ulem} 

\newcommand{\Tr}{\mathrm{Tr}}

\newcommand{\rr}{\mathbf{r}}
\newcommand{\rrp}{\mathbf{r}'}
\newcommand{\ax}{\mathbf{a}_x}
\newcommand{\ay}{\mathbf{a}_y}

\definecolor{jhcolor}{rgb}{0.0, 0.6, 0.6}  

\definecolor{ForestGreen}{RGB}{34,139,34}

\begin{document}

\title{A One-Particle Density Matrix Framework for Mode-Shell Correspondence:
Characterizing Topology in Amorphous Higher-Order Topological Insulators}

\author{Miguel F.\ Martínez}
\affiliation{Department of Physics, KTH Royal Institute of Technology, 106 91, Stockholm, Sweden}
\author{Lucien Jezequel}
\affiliation{Department of Physics, KTH Royal Institute of Technology, 106 91, Stockholm, Sweden}
\author{Jens H.\ Bardarson}
\affiliation{Department of Physics, KTH Royal Institute of Technology, 106 91, Stockholm, Sweden}
\author{Thomas Klein Kvorning}
\affiliation{Department of Physics, KTH Royal Institute of Technology, 106 91, Stockholm, Sweden}
\author{Julia D.\ Hannukainen}
\affiliation{T.C.M. Group, Cavendish Laboratory, J.J. Thomson Avenue, Cambridge CB3 0US, United Kingdom}

\begin{abstract}
We present a framework for characterizing higher-order topological phases directly from the one-particle density matrix, without any reference to an underlying Hamiltonian.
Our approach extends the mode–shell correspondence, originally formulated for single-particle Hamiltonians, to Gaussian states subject to chiral constraints.
In this correspondence,  the mode index counts topological boundary modes, while the shell index quantifies the bulk topology in a region surrounding the modes, providing a bulk-boundary diagnostic.
In one-dimensional topological insulators, the shell index reduces to the local chiral marker, recovering the winding number in the translation-invariant limit.
We apply the mode-shell correspondence to a $C_4$ symmetric higher-order topological insulator with a chiral constraint and show that a fractional shell index implies that the higher-order phase is intrinsic.
The one-particle density matrix is formulated in real space, so the mode-shell correspondence also applies to models without translation invariance.
By introducing structural disorder into the $C_4$-symmetric higher-order insulator, we show that the mode-shell correspondence remains a meaningful diagnostic in amorphous structures.
The mode-shell correspondence generalizes to interacting states with a gapped bulk spectrum in the one-particle density matrix, providing a practical and diverse route to characterize higher-order topology from the quantum state itself.
\end{abstract}

\maketitle

\section{Introduction}

While the topology of symmetry protected phases is frequently described using Hamiltonians, it is ultimately a property of a quantum state itself. 
Topological states exhibit global features that remain invariant under local unitary transformations that preserve the symmetries of the state.
Gaussian states protected by local internal symmetries are fully classified~\cite{schnyder08, kitaev09,Ryu2010, ludwig15} within the Altland–Zirnbauer scheme~\cite{cartan26,zirnbauer96,Altland97}.
In these phases, the bulk–boundary correspondence ensures that a nontrivial bulk topology produces robust gapless boundary modes.

Introducing nonlocal spatial symmetries, such as rotation, mirror, and inversion, extends the classification of Gaussian states to topological crystalline phases~\cite{Ando_2015,Neupert_2018}, and higher-order topological phases~\cite{Benalcazar2017quantized,langbehn2017reflection-symmetric, Geier2018,Schindler2018,Trifunovic2019,Trifunovic2021,Chaou2023,xie2021higher}. 
In higher-order phases, the symmetry protects gapless corner or hinge modes localized on boundary regions invariant under the relevant spatial symmetries~\cite{Geier2018,Schindler2018}.
Such phases can be intrinsic, where spatial symmetries enforce a nontrivial bulk topology that guarantees the existence of these modes regardless of lattice termination.
Extrinsic phases, in contrast, lack a bulk topology that alone ensures boundary modes; their presence depends on specific boundary terminations compatible with the bulk symmetries~\cite{Geier2018,Trifunovic2019,Trifunovic2021}.
Although crystalline symmetries play a central role in higher-order topology, such phases can also emerge in noncrystalline~\cite{Varjas2019}, and disordered matter~\cite{Agarwala2020,Tao2023,Peng_2022,Chaou2025}.

The topology of Gaussian quantum states is fully encoded in their one-particle density matrix, containing all single-particle correlations.
Unitary symmetries decompose the one-particle density matrix into exponentially localized blocks, each shaped by the full set of local symmetry constraints.
These blocks correspond to distinct topological phases of Gaussian states, and determine their classification---For a Gaussian state the one-particle density matrix is a projector onto the occupied bands, defining the Brillouin zone bundle of the corresponding topological class.
The topology of Altland–Zirnbauer classes is characterized by momentum space topological invariants such as Chern, and winding numbers, and more recently through many different types of real-space invariants~\cite{Prodan2010, Loring2010, Prodan2011,Bianco2011,Loring2015,Huang2018,Hofstetter2019,Loring2019, Hughes2019,jezequel2022,Hannukainen2022, Hannukainen2024,LocalityAHC,LocalMarker,caio2019topological, dOrnellas2022,kitaev2006anyons,loring2019spectral, loring2020spectral, schulz2021spectral, schulz2022invariants, schulz2023spectral,franca2024topological, stoiber2024spectral, doll2021skew, cerjan2023spectral, doll2024local, schulz2024topological,jezequel2025localizer}.
Many of these invariants are defined in terms of an underlying Hamiltonian, but there are local topological markers formulated directly using the one-particle density matrix to characterize the topology of the state itself~\cite{Bianco2011,Hannukainen2022, Hannukainen2024}.
The one-particle density matrix~\cite{Penrose1956,Koch_2001,Bera2015,bera2017one,Lezama2017,Kells2018,Hannukainen2022,Hannukainen2024} provides a framework for local topological markers that capture the topology of disordered states, such as amorphous matter~\cite{agarwala2017,Mansha2017,Xiao2017,mitchell_amorphous_2018,Bourne:2018jr,Poyhonen2018,minarelli_engineering_2019,chern_topological_2019,mano_application_2019,Costa:2019kc,marsal_topological_2020,Sahlberg2020,ivaki_criticality_2020,Agarwala2020,wang_structural-disorder-induced_2021,focassio_structural_2021,Mitchell2021,spring_amorphous_2021,wang_structural_2022,Peng_2022,uria-alvarez_deep_2022,spillage_2022,Cassella2023,Grushin2023,Corbae2023,Manna2024,marsal_obstructed_2022, uria2024amorphization}, as well as certain physically relevant interacting states, including midspectrum many-body localized states.

Higher-order topology has been characterized using a variety of methods, including, for example, nested Wilson loops, quantized multipole moments, chiral multipole invariants, and symmetry-based indicators~\cite{alexandradinata2014Wilson, Benalcazar2017quantized, Benalcazar2017electric}.  
Several approaches have also been developed to define invariants for higher-order topological insulators that lack translation invariance~\cite{Agarwala2020,Tao2023,Peng_2022,zijderveld2025scatteringtheoryhigherorder,CerjanHOTI}.  

The mode–shell correspondence~\cite{Jezequel2024,Jezequel2025} offers a method  for characterizing topology in states subject to a chiral constraint.
This formalism establishes a correspondence between two equivalent topological indices: the mode index and the shell index.
The mode index counts the chiral zero modes of a single-particle Hamiltonian, while the shell index is computed in a region surrounding these modes encoding the bulk topology.
The mode-shell correspondence was originally developed as a characteristic of topology of single-particle Hamiltonians in any dimension.
In this work, we reformulate the mode–shell correspondence in terms of the one-particle density matrix, enabling the characterization of higher-order topological phases directly from the many-body state.
We derive the mode and shell indices directly from the one-particle density matrix for states subject to a chiral constraint, resulting in a formulation of the mode-shell correspondence for Gaussian states.
We explain how the shell index in one dimension is equivalent to the local chiral marker~\cite{Hannukainen2022}, a real space topological bulk invariant that reduces to the chiral winding number in the translation-invariant limit.
We then apply this framework to a $C_4$ rotationally symmetric higher-order topological insulator in two dimensions, demonstrating that the mode–shell correspondence distinguishes between intrinsic and extrinsic phases with an odd number of corner modes.
The one-particle formalism extends to disordered structures, and we use the mode-shell correspondence to characterize the topology of an amorphous higher order topological insulator, with enforced $C_4$ symmetry.

\section{The one-particle density matrix and the topology of states}

The Bogoliubov–de Gennes one-particle density matrix provides a framework for classifying the topology of Gaussian states. 
For a many-body state $\lvert \Psi \rangle$, the one-particle density matrix has the block form  
\begin{equation}
\rho =
\begin{pmatrix}
\tilde{\rho} & \kappa \\
\kappa^\dagger & 1 - \tilde{\rho}^* 
\end{pmatrix},\label{eq:OPDM}
\end{equation}
with $\tilde{\rho}_{ij} = \langle \Psi| c_i^\dagger c_j |\Psi \rangle$ and $\kappa_{ij} = \langle \Psi| c^\dagger_i c^\dagger_j |\Psi \rangle$, where $c_j^\dagger$ and $c_j$ are fermion creation and annihilation operators~\cite{Bera2015,bera2017one,Kells2018}.
The eigenvectors of $\rho$ define natural orbitals with occupations $0 \le n_\alpha \le 1$, which are either zero or one for Gaussian states~\cite{Penrose1956,Koch_2001,Bera2015,bera2017one,Lezama2017}. 
The topological classification applies to Gaussian states whose one-particle density matrix is exponentially localized in the bulk---The one-particle density matrix is exponentially localized in real space if $\braket{\rr|\rho|\rrp}\sim e^{-|\rr-\rrp|/\xi}$, where $\ket{\rr}, \ket{\rrp}$ are single-particle states localized at positions $\rr,\rrp$ and $\xi$ is the correlation length.
In the translation invariant limit, $\rho$ reduces to a projector onto the occupied single-particle orbitals---the image of $\rho$ constitutes a vector space at each momenta, forming a vector bundle over the Brillouin zone.
Unitary symmetries block-diagonalize $\rho$ into symmetry invariant subspaces, and within each block local unitary or antiunitary symmetries impose further constraints. 
It is these blocks---or equivalently the associated vector bundle---that are classified into symmetry classes, determining the possible topological phases~\cite{kitaev09,schnyder08,Ryu2010}.
In this work, we focus on Gaussian states, but the topological classification based on the one-particle density matrix also applies to interacting systems, provided the occupation spectrum $\{ n_\alpha\}$ remains gapped.
If the one-particle density matrix of an interacting state is gapped, it can be adiabatically flattened into a projector, while remaining exponentially localized in the bulk~\cite{Hannukainen2024}.
This means that the topology of the band flattened one-particle density matrix, $\varrho = \big[(2\rho - 1)/|2\rho - 1| + 1 \big]/2$, defines the topology of the interacting state.

\section{The mode-shell correspondence}

The mode-shell correspondence defines the general equivalence between the chiral zero modes of a single-particle Hamiltonian, enumerated by the mode index $\mathcal{I}_{\rm_{mode}}$, and a bulk property, defined as the shell index \(\mathcal{I}_{\rm shell}\), of the same Hamiltonian. 
This equivalence emerges by rewriting \(\mathcal{I}_{\rm mode}\) as an expression involving only bulk operators. 
In this work, we extend the mode-shell correspondence to a classification of quantum states by expressing it in terms of the one-particle density matrix. 

The mode index, defined as $\mathcal{I}_{\rm mode} = n_+ - n_-$, counts the net difference between the number of boundary zero modes of positive $n_+$ and negative chirality $n_-$ in an infinite system~\cite{kaneLubenski,MarceloTango}.
A given mode index for a single-particle Hamiltonian implies correlations between the topological modes at different edges in the ground state of that Hamiltonian.
Such correlations are encoded in the one-particle density matrix.
To make the connection between the mode index and the one-particle density matrix explicit, we define a mode index from the restriction of the one-particle density matrix to a region enclosing the chiral edge modes.
For clarity, we focus our discussion on Gaussian one-dimensional topological states with a chiral constraint. 
However, the methods and results extend to higher dimensions, higher-order topological insulators, and to interacting states; as long as the one-particle density matrix is gapped and band flattened.
\begin{figure}[t!]
    \centering
\includegraphics[width=0.95\linewidth]{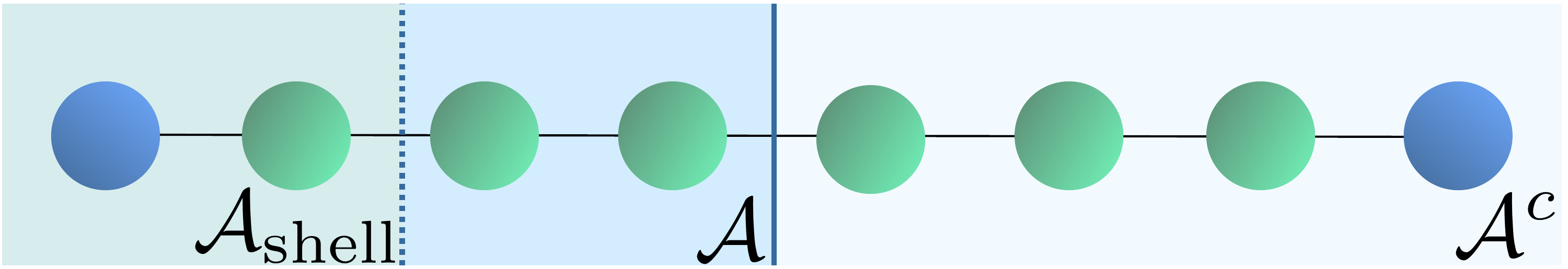}
    \caption{Example of the choice of regions $\mathcal{A}$, its complement $\mathcal{A}^c$, and $\mathcal{A_{\rm shell}}\subset \mathcal{A}$
    in a one-dimensional chain with the topological zero modes, depicted in blue, at the two ends of the chain.
    The solid line represents the boundary between $\mathcal{A}$ and $\mathcal{A}^c$, and the dotted line represents the shell, the boundary of $\mathcal{A_{\rm shell}}$.
    \label{Fig:chain}
    }
\end{figure}

We consider a one-dimensional lattice with a spatial region $\mathcal{A}$ containing one of the ends of the chain, as exemplified in Fig.~\ref{Fig:chain}. 
The fermionic ground state on the chain determines a one-particle density matrix $\rho$.
Restricting $\rho$ to $\mathcal{A}$ defines the restricted one-particle matrix $\rho_{\mathcal{A}}$, a block $\rho$ with matrix elements $\rho_{ij}$ for $i,j \in \mathcal{A}$.
This block coincides with the reduced one-particle density matrix obtained by tracing out the complement of $\mathcal{A}$.
We assume $\rho$, and consequently $\rho_\mathcal{A}$, to satisfy a chiral constraint $\{\rho, C\} = C$ coming from the combination of time-reversal symmetry and the inherent particle-hole structure of the one-particle density matrix [Eq.~(\ref{eq:OPDM})]. 
The operator $C$ is a local single-particle operator with eigenvalues $\pm1$.
The eigenvalues $\lambda$ of $\rho_{\mathcal{A}}$ reflect the occupations of single-particle modes localized within the region $\mathcal{A}$: $\lambda=1$ corresponds to fully occupied modes and $\lambda=0$ to empty modes.
Eigenvalues taking intermediate values represent modes that are partially occupied, which is only possible if they exhibit correlations extending beyond the boundary of $\mathcal{A}$.
The eigenvalues exactly equal to $\lambda= 1/2$ signal modes that are maximally entangled between $\mathcal{A}$ and its complement. 
These eigenvalues may have a topological origin, as topological zero modes hybridize between the different ends of the chain, or a purely accidental one, for example, a single-particle bulk mode localized on a region split by the boundary of $\mathcal{A}$.

As a first step toward identifying the topological edge correlations, we focus on eigenstates of $\rho_\mathcal{A}$ that are correlated across the boundary of $\mathcal{A}$.
These eigenstates are singled out by the sum
\begin{align}
\mathcal{I}= \sum_{\lambda} 4\lambda(1-\lambda),
\label{eq_intermediate_sum}
\end{align}
which equals to one for $\lambda=1/2$, and vanishes for $\lambda=0,1$, thereby excluding modes confined entirely to $\mathcal{A}$ or its complement.
The sum $\mathcal{I}$ accounts for all modes with support across the boundary of $\mathcal{A}$, meaning all $0 < \lambda<1$, and not only the chiral edge modes pinned at $\lambda=1/2$. 
As a result,  Eq.~\eqref{eq_intermediate_sum} over estimates the mode index by including possible contributions from trivial correlations between the edges of the chain, and local correlations across the boundary of $\mathcal{A}$.

To isolate the topological edge correlations, we introduce the overlap 
\begin{align}
\mathcal{O}_\lambda =\bra{\lambda}\theta C\ket{\lambda},
\label{eq_overlap}
\end{align}
multiplying each term in the sum in Eq.~\eqref{eq_intermediate_sum}, where $\ket{\lambda}$ are the eigenstates of the restricted one-particle density matrix with corresponding eigenvalues $\lambda$.
The operator $\theta$ acts as a filter projecting into a subregion $\mathcal{A}_{\rm shell} \subset \mathcal{A}$ containing the chiral edge modes (see Fig.~(\ref{Fig:chain})).
The operator $\theta$ is therefore diagonal in the position basis, where the diagonal entries are chosen as any smooth function of position that equals to one within $\mathcal{A}_{\rm{shell}}$ and decays to zero outside.
We refer to the boundary of $\mathcal{A}_{\rm shell}$, identified by $\nabla \theta \neq 0$, as the shell.
The overlap $\mathcal{O}_\lambda$ can only be nonzero for modes that have support inside $\mathcal{A}_{\rm shell}$.
This means that $\mathcal{O}_\lambda$ is zero for any mode that is locally correlated across the boundary of $\mathcal{A}$, as it has no support in the shell region.
Bulk modes with support both in $\mathcal{A}_{\rm shell}$ and in $\mathcal{A}^{\rm c}$, the complement of region $\mathcal{A}$, can have a nonzero $\mathcal{O}_\lambda$.
The distance between the shell and the boundary of $\mathcal{A}$ must therefore be larger than the correlation length in order to exclude such trivial contributions.
Any other local modes with support in $\mathcal{A}_{\rm shell}$ also have a vanishing $\mathcal{O}_\lambda$, since the locality of the chiral symmetry operator enforces that the chiral pairs $\ket{\lambda}$ and $C\ket{\lambda}$ are locally orthogonal.  
Nonlocal modes resulting in trivial correlations between the two ends of the chain do not yield a contribution to the weighted sum, even though their individual overlaps are nonzero.
In this case, the restricted one-particle density matrix has degenerate eigenstates with $\lambda=1/2$ that appear in symmetric and antisymmetric combinations of the two ends.  
The chiral symmetry operator $C$ flips the relative sign between the two chiral partners, so that one mode has an overlap $\mathcal{O}_\lambda=+1$, and its chiral partner an overlap $\mathcal{O}_\lambda=-1$. 
As both eigenvalues are summed with opposite-sign weights, the two modes cancel, so trivial edge correlations do not contribute to the index.

The chiral edge modes do contribute to the sum as they have weight in $\mathcal{A}_{\rm shell}$ and their overlap in Eq.~\eqref{eq_overlap} is equal to $\pm1/2$ up to exponential corrections.
Chiral edge modes are in equal weight superposition between the two ends of the chain, with an opposite relative sign between the chiral partners.
As $\theta$ filters out one of the ends, the only contribution from the overlap comes from the end enclosed by $\mathcal{A}$, where half of the weight of the two chiral partners is localized.
Since the localization at the ends is exponential, the deviations from  $\pm1/2$ for different choices of $\mathcal{A_{\rm shell}}$ are exponentially suppressed, as long as $\mathcal{A_{\rm shell}}$ encloses a sufficiently large region compared to the localization length of the modes.
With both local correlations and trivial edge-to-edge correlations excluded, only the chiral edge modes remain.
By combining the sum in Eq.~\eqref{eq_intermediate_sum}, and the overlap in Eq.~\eqref{eq_overlap} we define the mode index to be
\begin{align}
\mathcal{I}_{\rm_{mode}} = \sum_{\lambda}4\lambda(1 - \lambda)\langle \lambda|\theta C |\lambda \rangle,
\label{eq:mode_index_sum}
\end{align}
which, in operator form, is expressed as
\begin{align}
\mathcal{I}_{\rm_{mode}}=4\Tr\left((\rho_{\mathcal{A}} - \rho_{\mathcal{A}}^2)\theta C\right).
\label{eq:mode_index}
\end{align}
The mode index sums the contributions from each chiral edge state, resulting in $\mathcal{I}_{\rm{mode}} = 1$ for the one dimensional chain. 
By using the linearity and the cyclic property of the trace operation and the constraint $\{\rho_{\mathcal{A}},C\}=C$, the first term in the mode index in Eq.~\eqref{eq:mode_index} can be rewritten as
\begin{align}
4\Tr(C\rho_{\mathcal A}\theta)&=2\Tr(C\rho_{\mathcal A}\theta)+\Tr(C\theta)+\Tr([C\rho_{\mathcal A},\theta]),
\label{eq:mode_deriv_1}
\end{align}
where the last term, being a trace of a commutator, is zero.
The second term in Eq.~\eqref{eq:mode_index} is similarly expanded as
\begin{align}
-4\Tr(C\rho_{\mathcal A}^2\theta)
&=-2\Tr\!\big((C-\rho_{\mathcal A}C)\rho_{\mathcal A}\theta\big)-2\Tr(C\rho_{\mathcal A}^2\theta) \nonumber\\
&=-2\Tr(C\rho_{\mathcal A}\theta)+2\Tr\!\big(C\rho_{\mathcal A}[\theta,\rho_{\mathcal A}]\big).
\label{eq:mode_deriv_2}
\end{align}
Combining Eqs.~\eqref{eq:mode_deriv_1} and \eqref{eq:mode_deriv_2} yields the expression:
\begin{equation}
4\Tr\left(C(\rho_{\mathcal{A}} - \rho_{\mathcal{A}}^2)\theta\right)
=  \Tr(C\theta)- 2\Tr(C\rho_{\mathcal{A}}[\rho_{\mathcal{A}}, \theta]).
\label{eq:expand_mode_index}
\end{equation}
A nonzero contribution from $\Tr(C\theta)$ would require an imbalance between positive and negative modes, corresponding to chiral constraints that are generally artificial and rarely encountered in typical condensed matter systems.
In our case, the first term always vanishes because the chiral constraint stems from the inherent particle-hole constraint of the Bogoliubov-de Gennes one-particle density matrix.
The remaining term in Eq.~\eqref{eq:expand_mode_index} is only nonzero on the shell---the commutator $[\rho_\mathcal{A}, \theta]$ vanishes away from the shell since $\theta$ is constant in these regions.
This means that the mode index in Eq.~(\ref{eq:mode_index}) can be expressed in terms of an index which is only nonzero on the shell, defining the shell index:
\begin{equation} 
\label{eq:shell_index}
    \mathcal{I}_\text{shell} = -2\Tr(C\rho_\mathcal{A}[\rho_\mathcal{A}, \theta]).
\end{equation}  
When $\theta(\mathbf{x})$ varies smoothly over a length scale larger than the correlation length $\xi$ of $\rho_{\mathcal{A}}$,  the commutator $[\rho_\mathcal{A}, \theta]$ becomes small and can be expanded in a gradient expansion, as shown in detail in appendix~\ref{app:gradient_expansion}.
The commutator $[\rho_\mathcal{A}, \theta] \approx \sum_i [\rho_\mathcal{A}, X_i] \, \partial_{x_i} \theta$, to leading order in $[\rho_\mathcal{A}, X_i]$, where $X_i$ are position operators in general spatial dimensions.
This approximation is controlled by the exponential localization of $\rho_\mathcal{A}$, $\langle x|\rho_\mathcal{A}|x'\rangle \sim e^{-|x-x'|/\xi}$, and the leading correction scale as $\xi^2\partial_{x_i}\partial_{x_j}\theta$. 
This results in the alternative expression
\begin{align}
\label{eq:shell_index_two}
\mathcal{I}_\text{shell} = -2\sum_i \Tr\left(C\rho_\mathcal{A}[\rho_\mathcal{A}, X_i] \, \partial_{x_i} \theta\right),
\end{align}
for the shell invariant, explicitly showing that the invariant is localized in the region where $\theta$ changes.

In one dimension, the shell index is, up to a sign, equivalent to the local chiral marker of the restricted one-particle density matrix (see Appendix~\ref{app:chiral_marker} for a derivation), 
\begin{equation}
\nu(x)=-2\text{tr}\langle x|\rho_\mathcal{A} C X\rho_\mathcal{A}|x\rangle,
\label{eq:chiral_marker}
\end{equation}
a real-space $\mathbb{Z}$ invariant equal to the chiral winding number in translation-invariant models~\cite{kaneLubenski, Hannukainen2022}.
The lower case trace tr in Eq.~\eqref{eq:chiral_marker} represents a trace over any internal degrees of freedom.
$\theta$ is diagonal in position space, so $\mathcal{I}_{\text{shell}} = \sum_x \partial_x \theta(x)\nu(x)$, where we have used the cyclic property of the trace, translation invariance, and the chiral constraint in Eq.~\eqref{eq:shell_index_two} to change the order of the operators.
Since $\theta$ interpolates from zero to one across the chain,  $\sum_x \partial_x \theta(x) = \pm 1$, by discrete partial integration.
In the bulk region, where the shell is located and the chiral marker is constant, this yields
\begin{equation}
    \mathcal{I}_{\text{shell}} = \nu \sum_x \partial_x \theta(x) = \pm \nu.
\end{equation}
The sign is determined by which end of the chain is chosen as the restricted region $\mathcal{A}$.
In one dimension, the shell is a codimension-one object and therefore consists of an isolated point.
Since the bulk itself is one-dimensional, a codimension-one shell reduces to an isolated point that exists entirely within the bulk, away from the physical boundary, which is why the shell index is equal to the chiral winding number, a bulk invariant.
This argument does not carry over to higher dimensions as the shell is no longer purely contained in the bulk and $\mathcal{I}_{\rm shell}$ is therefore not a bulk invariant.
\begin{figure}[t!]
    \centering
\includegraphics[width=1\linewidth]{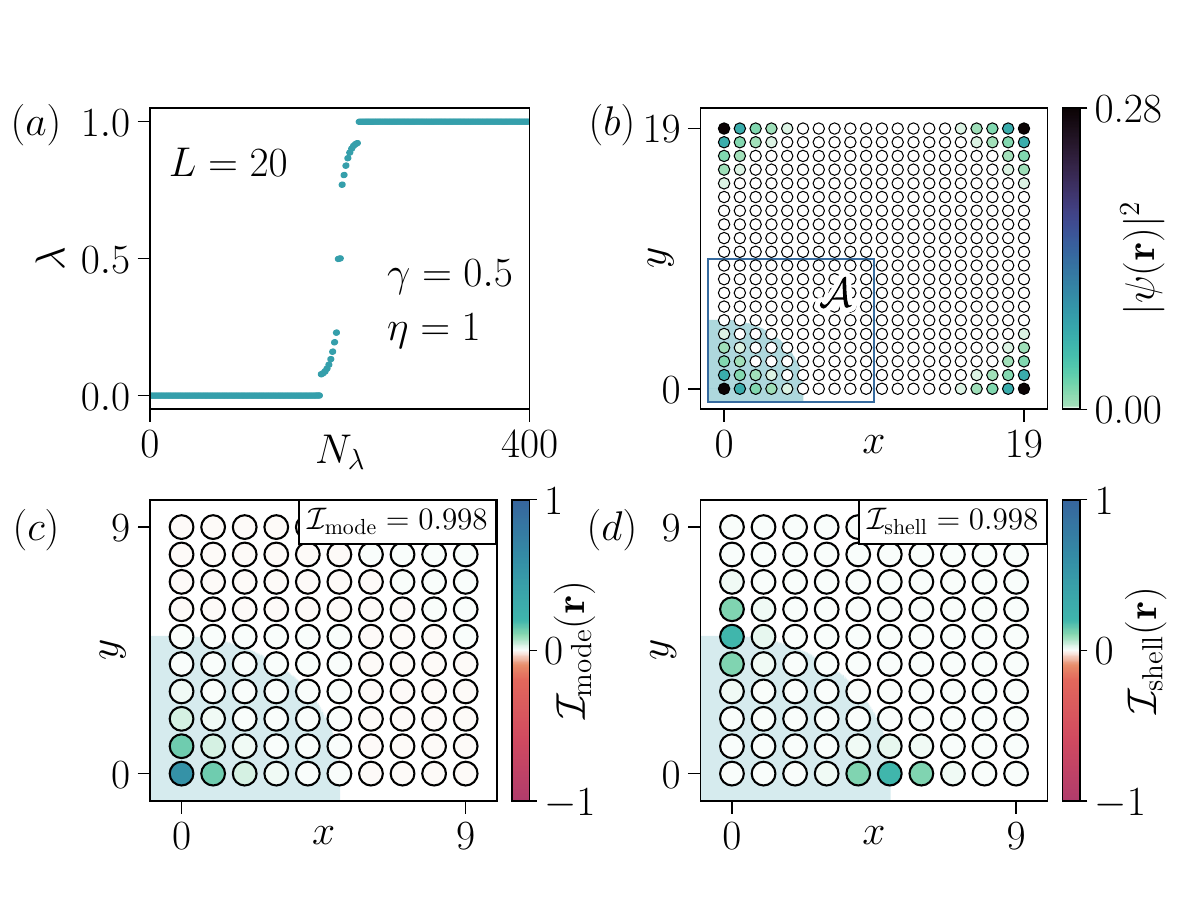}
    \caption{(a) Spectrum of the restricted one-particle density matrix $\rho_\mathcal{A}$ for the ground state of the BBH model at half filling, with parameters  $\gamma=0.5, \eta=1$ and defined on a square lattice of linear size $L=20$.
    $\lambda$ denotes the eigenvalues and $N_\lambda$ the eigenvector number.
    (b) Combined density of states of the four zero modes in the energy spectrum of the BBH Hamiltonian.
    The region $\mathcal{A}$ contains the lattice sites with $x<L/2$ and $y<L/2$, enclosed by the blue box, and the region $\mathcal{A}_{\rm shell}$ is shaded in light blue.
    (c) Mode index density, defined in Eq.~(\ref{eq:mode_density}), within the region $\mathcal{A}$. 
    The box in the upper right corner indicates the total mode index, computed by summing the mode index density over real space.
    (d) Shell index density, defined in Eq.~(\ref{eq:shell_density}), within the region $\mathcal{A}$.
    The box in the upper right corner indicates the total shell index, computed by summing the shell index density over real space.
     The region $\mathcal{A}_{\rm shell}$ is shaded in light blue in (c) and (d).
    \label{Fig:CrystallineHoti_mode}
    }
\end{figure}

The mode-shell correspondence formalism presented in this section is general and applies to chirally symmetric models in arbitrary dimensions, provided regions $\mathcal{A}$ and $\mathcal{A}_\text{shell} \subset \mathcal{A}$ are sufficiently large to isolate a topological mode. 
While the discussion so far has focused on one-dimensional topological insulators with end-localized modes, the same framework extends to two-dimensional higher-order topological insulators, where topological modes appear at the corners and their contributions to the mode index are constrained by crystalline symmetries.

\section{The mode-shell correspondence in higher-order topological insulators}

We consider the Benalcazar-Bernevig-Hughes (BBH) model~\cite{Benalcazar2017quantized,Benalcazar2017electric} to exemplify how the mode-shell correspondence applies to higher-order topological insulators.
The BBH model describes a two dimensional higher-order topological insulator on a square lattice protected by a combination of time-reversal symmetry, which implies a chiral constraint, and $C_4$ rotational symmetry, and hosts four zero energy topological corner modes.
The Bogoliubov–de Gennes Hamiltonian of the BBH model in real space is
\begin{equation}
\begin{split}
\label{eq:BBH_hamiltonian}
    h_{
    \rr \rrp} = -\gamma(\sigma_z + & \sigma_y \tau_y)\delta_{\rr\rrp} - \dfrac{\eta}{2}(\sigma_z  + i \sigma_y \tau_z) \delta_{\rrp,\rr+\ax} \\ 
    - & \dfrac{\eta}{2} (\sigma_y  \tau_y + i \sigma_y \tau_x) \delta_{\rrp,\rr+\ay} ,
\end{split}
\end{equation}
where $\sigma_j$ and $\tau_j$, with $j\in(x,y,z)$, are Pauli matrices acting on the Bogoliubov-de Gennes space and on the orbital degrees of freedom within a unit cell, respectively.
The vectors $\ax$ and $\ay$ connect the adjacent sites $\rr, \rrp$ in the $x,y$ directions of the square lattice, and take the form $\ax = (1,0)$ and $\ay = (0,1)$ when the lattice constant is set to unity.
The Hamiltonian $h_{\rr \rrp}$ has a chiral constraint $C=\sigma_x$.
The relative size of the hopping parameters $\gamma$ and $\eta$ determines the topological phase: Equation \eqref{eq:BBH_hamiltonian} describes an insulator in a high-order topological phase for $|\gamma/\eta|< 1$, with a phase transition at $|\gamma/\eta|=1$. 

We consider the ground state of the BBH model with $\gamma=0.5$ and $\eta=1$ at half filling on a square lattice with linear size $L=20$, and define the restricted region $\mathcal{A}$ to be a square of linear size $L_{\mathcal{A}}=10$, taken as the lower left quarter of the full lattice.
The spectrum of the restricted density matrix $\rho_\mathcal{A}$ shown in Fig.~\ref{Fig:CrystallineHoti_mode}(a), exhibits four eigenvalues pinned at  $\lambda=1/2$, corresponding to chiral corner modes, together with additional eigenvalues in the interval $0<\lambda<1$ that originate from trivial short range correlations across the boundary of $\mathcal{A}$.
To isolate the contribution of the chiral corner modes, we introduce the function
\begin{equation}
\label{eq:theta}
\theta(x,y)=1-\left[1+e^{-c\left(\sqrt{x^2+y^2}-\omega\right)}\right]^{-1},
\end{equation}
which projects onto a subregion $\mathcal{A}_{\rm{shell}}\in \mathcal{A}$, depicted by the shaded region in Fig.~\ref{Fig:CrystallineHoti_mode}(b).
Here, $w=L_{\mathcal{A}}/2$ determines the position of the shell and $c=L_{\mathcal{A}}/2$ controls the sharpness of the transition from $\theta=1$ to $\theta=0$.

Since both the mode and shell indices consist of an operator trace over the full Hilbert space, they admit a representation in terms of densities, where the indices are the real space sum of different local contributions.
This results in $\mathcal{I}_{\rm mode}=\sum_\rr\mathcal{I}_{\rm mode}(\rr)$ and $\mathcal{I}_{\rm shell}=\sum_\rr\mathcal{I}_{\rm shell}(\rr)$, with 
\begin{equation}
    \label{eq:mode_density}
    \mathcal{I}_{\rm mode}(\rr)=4\text{tr}\bra{\rr}(\rho_\mathcal{A}-\rho_\mathcal{A}^2)C\theta \ket{\rr},
\end{equation}
\begin{equation}
\label{eq:shell_density}
    \mathcal{I}_{\rm shell}(\rr)=-2\text{tr}\bra{\rr}C\rho_\mathcal{A}[\rho_\mathcal{A}, \theta]\ket{\rr}.
\end{equation}
These densities, together with the total value of both indices, are shown in  Figs.~\ref{Fig:CrystallineHoti_mode}(c) and~\ref{Fig:CrystallineHoti_mode}(d) .
The numerical results show that the mode index in Eq.~\eqref{eq:mode_index_sum} and the shell index in Eq.~\eqref{eq:shell_index} both approach one, reflecting the presence of a topological corner mode. 
The mode index density is located around the corner of $\mathcal{A}$, while the shell index density contains two equal contributions, one from each intersection between the shell and the lattice boundary.
The absence of bulk contributions to the shell index can be understood by considering one-dimensional slices perpendicular to the edges of the square lattice.
For each slice, one can evaluate a chiral winding number that plays the role of a weak topological bulk invariant which is equivalent to the shell index of the slice.
If the shell index is nonzero at any point in the bulk, the corresponding chiral winding number must also be nonzero, which in turn implies the existence of a weak topological insulator with topological zero-modes localized at the ends of the slice, along the system boundary~\cite{yang_experimental_2022,StrongsideWeakTI,Disorweakandstrong,AperiodicWTI,PredictionWeak}. 
In the BBH model and higher-order topological insulators in general, such chiral edge modes are forbidden by their defining topology, with gapless states appearing exclusively at the corners.
Thus, all bulk contributions to the shell index vanish: The shell index receives support solely from the intersections of the shell with the boundary, each contributing with a factor of $\mathcal{I}_{\rm{mode}}/2$ enforced by the $C_4$ symmetry.

The BBH model realizes an intrinsic higher-order topological insulator: Its corner modes are robust and cannot be removed by changing the edge termination or by attaching additional chiral-symmetry-preserving one-dimensional chains along the edge.
The intrinsic nature of the BBH model can be deduced from the parity of the mode index.
Whenever the mode index is odd, each edge-shell  intersection contributes a half-integer value to the shell index. 
These half-quantized contributions cannot be canceled by changing the edge termination; adding a one dimensional chain would only change the shell index by integer units.
This means that an odd mode index can be used as an indicator that the higher-order topological insulator in question is intrinsic.
In contrast, an even mode index does not distinguish intrinsic from extrinsic higher-order topological insulators,  since the contributions from the shell index on each edge are integer valued and potentially trivialized by edge modifications.

\section{The BBH model on an amorphous structure}

The mode and shell indices, Eq.~\eqref{eq:mode_index_sum} and Eq.~\eqref{eq:shell_index}, are expressed in terms of the one-particle density matrix,  which means that the mode-shell correspondence is applied directly in real space to characterize the topology of states, irrespectively of translation invariance.
Without translation invariance, the one-particle density matrix  is no longer block-diagonal in momentum space, and the associated vector-bundle structure over the Brillouin zone is lost. 
As a result, the topological classification based on families of vector bundles no longer applies. 
However, the single-particle states retain a translation-invariant long-wavelength limit, in which the coarse-grained one-particle density matrix  $\tilde{\rho}$ becomes effectively translation invariant.
In this limit, the mode and shell indices are well defined as topological characteristics of the state.
In practice, the lack of translation invariance induces a change in the localization length of the topological modes, resulting in a spreading of the local contributions to the mode index.
As long as the shell region is large enough to account for this spreading, the mode index will approach the quantized value characteristic of the crystalline limit, and calculations are performed using $\rho$ instead of $\tilde{\rho}$.
In the case of the shell index, the change in the correlation length of $\rho$ due to the lack of translation invariance may result in nonzero local bulk contributions to $\mathcal{I}_{\rm shell}$.
The average of these contributions in regions much larger than the correlation length of $\rho$ vanishes---Upon coarse graining, translation invariance is recovered and the bulk contributions to the shell index must be zero. 

To demonstrate the mode-shell correspondence in a disordered structure, we apply the BBH model to a $C_4$-symmetric amorphous point set and characterize the topology of its ground state.
In order to generate a $C_4$-symmetric amorphous square point set of linear size $L$, we consider a smaller square lattice with linear size $\tilde{L}=L/2$.
The smaller lattice is made into an amorphous structure by drawing each site from a normal distribution with a mean given by the lattice coordinates of the sites, and with a standard deviation $w$.
We denote this amorphous structure as the generating set.
The parameter $w$ controls the deviation from the crystalline limit and is referred to as the \textit{amorphicity} of the model.
The final amorphous point set consists of four copies of the generating set arranged in a square.
To enforce a $C_4$ symmetry, the copy in the top right corner is left unchanged, while the remaining three are rotated by $\pi/2$, $ \pi$, and $3\pi/2$ around their own center axis, and placed in the top left, bottom left, and bottom right corners, respectively.
While the $C_4$ symmetry of the amorphous point set is needed for the zero modes to be protected against chiral-symmetry-preserving perturbations, it is not required for the application of the mode-shell correspondence.

The BBH Hamiltonian defined on the amorphous structure described above  is~\cite{Agarwala2020}
\begin{equation}
\begin{split}
    h_{
    \rr \rrp} = -&\gamma(\sigma_z +  \sigma_y \tau_y)\delta_{\rr\rrp} - \dfrac{\eta_{\rrp\rr}}{2}(\sigma_z + i \sigma_y \tau_z) \cos \phi\\ 
    - & \dfrac{\eta_{\rrp\rr}}{2} (\sigma_y  \tau_y + i \sigma_y \tau_x) \sin \phi,
    \end{split}
    \label{eq:AmorphousBBH}
\end{equation}
where $\phi$ is the angle between the vector $\rrp-\rr$, which connects neighboring sites, and the $x$ axis.
The parameter $\eta_{\rrp \rr}$ depends on the relative position between sites as
\begin{equation}
    \eta_{\rrp \rr}=\eta \Theta(\vert \rrp -\rr\vert - r_0)e^{-\vert \rrp -\rr\vert+1} \quad \rm{for} \quad \rrp\neq\rr,
\end{equation}
and vanishes for $\rrp=\rr$.
The cutoff distance $r_0$ determines the maximum distance between sites connected by the hopping.
We set $r_0=1.3$, and the lattice constant of the reference crystalline lattice as $a=1$.
\begin{figure}[t!]
    \centering
\includegraphics[width=1\linewidth]{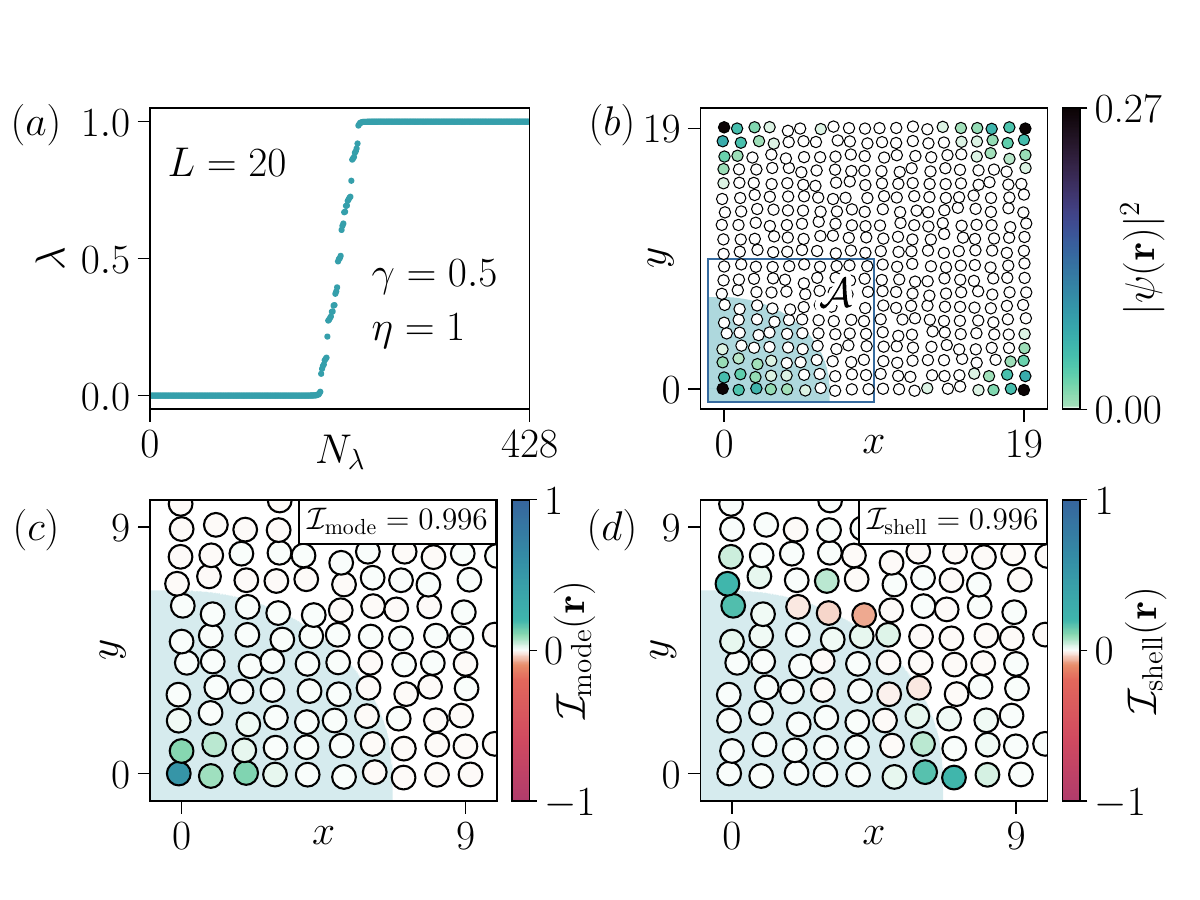}
    \caption{(a) Spectrum of the restricted one-particle density matrix $\rho_\mathcal{A}$ for the ground state of the BBH model at half filling, with parameters  $\gamma=0.5, \eta=1$ and defined on an amorphous structure of linear size $L=20$, and amorphicity $w=0.1$.
    $\lambda$ denotes the eigenvalues and $N_\lambda$ the eigenvector number.
    (b) Combined density of states of the four zero modes in the energy spectrum of the BBH Hamiltonian.
    The region $\mathcal{A}$ contains the lattice sites with $x<L/2$ and $y<L/2$, enclosed by the blue box, and the region $\mathcal{A}_{\rm shell}$ is shaded in light blue.
    (c) Mode index density, defined in Eq.~(\ref{eq:mode_density}), within the region $\mathcal{A}$. 
    The box in the upper right corner indicates the total mode index, computed by summing the mode density over real space.
    (d) Shell index density, defined in Eq.~(\ref{eq:shell_density}), within the region $\mathcal{A}$. 
    The box in the upper right corner indicates the total shell index, computed by summing the shell index density over real space.
    The region $\mathcal{A}_{\rm shell}$ is shaded in light blue in (c) and (d).
    \label{Fig:AmorphousHoti_mode}
    }
\end{figure}

We evaluate the mode and shell indices for the ground state at half filling of the BBH model   [Eq~\eqref{eq:AmorphousBBH}] in an amorphous structure with $L=20$, amorphicity $w=0.1$,  $\gamma=0.5$, and $ \eta=1$, for which the spectrum of the restricted one-particle density matrix is shown in Fig.~\ref{Fig:AmorphousHoti_mode}(a) and the density of states of the four zero modes in the spectrum is displayed in Fig.~\ref{Fig:AmorphousHoti_mode}(b).
We choose the restricted region $\mathcal{A}$ as the lower left quarter of the amorphous structure, and specify the cut off function $\theta$ in Eq.~\eqref{eq:theta}.
In this case, we choose the parameters  $c=2L_{\mathcal{A}}/3$ and $\omega=L_{\mathcal{A}}/2$, to account for the increase in the localization length of the zero modes due to amorphicity.

The structural disorder modifies the localization length of the eigenmodes of the full one-particle density matrix, leading to more modes developing short-range correlations across the boundary of $\mathcal{A}$ than in the crystalline limit.
As a consequence, more eigenvalues of the restricted spectrum appear in the interval $0<\lambda<1$, Fig.~\ref{Fig:AmorphousHoti_mode}(a), compared to the crystalline limit.
Figures \ref{Fig:AmorphousHoti_mode}(c) and \ref{Fig:AmorphousHoti_mode}(d)  show the mode and shell index densities for a single realization of the amorphous structure; the shell index contains contributions from the edge and the bulk of the shell, contrary to the crystalline limit, which only hosts edge contributions.
The presence of both negative and positive bulk contributions points to the cancellation between the different bulk contributions to the shell index when averaged over sufficiently large distances.

\section{Discussion}

We derived the mode–shell correspondence in terms of the one-particle density matrix and showed that the shell index is equivalent to the local chiral marker in one-dimensional chains with a chiral constraint.
The formalism developed in this work is directly applicable to many-body states, without a reference to a parent Hamiltonian, reflecting that topology is a property of the state itself.
In addition, the one-particle density matrix enables a real space characterization of topology, independent of translation invariance.

We used the mode-shell correspondence to characterize the topology of $C_4$ symmetric higher-order topological phases with a chiral constraint.
While the mode and shell indices are not bulk invariants, they are still relevant for characterizing higher-order topological insulators. 
We argued that a half-integer boundary contribution to the shell index identifies an intrinsic higher-order phase, whereas an integer contribution to the boundary is not enough to differentiate between intrinsic and extrinsic phases.
Our examples focus on two-dimensional, $C_4$-symmetric models, but the construction is general. It applies to higher dimensions and other crystalline symmetries, such as higher-order rotations or mirror symmetry, that map different parts of the shell onto one another. Consequently, their contributions must be equal, leading to a fractional quantization and therefore intrinsic topology.

By computing the mode and shell indices for higher-order topological insulators in both crystalline and amorphous structures, we demonstrated that their correspondence provides a tool to diagnose topological states with crystalline symmetries, which must be enforced by hand in amorphous models.
The one-particle density matrix is central to this formulation: It characterizes topology directly in real space, and remains well defined in both crystalline and disordered settings.
An additional benefit of the formalism is that it extends to interacting states within a parameter regime where the one-particle density matrix of the interacting state is gapped~\cite{Hannukainen2024}.
As long as its gap remains open, the one-particle density matrix can be band-flattened into a projector, which means that the mode-shell correspondence with respect to the band-flattened density matrix defines the topology of the interacting state.
Although this parameter regime is limited, it includes physically relevant states in condensed matter, such as the ground states of weakly interacting Hamiltonians and many-body localized eigenstates~\cite{Bera2015, bera2017one, Lezama2017}.

\section{Acknowledgements}
We thank Adam Yanis Chaou for useful discussions.
This work received financial support from the Swedish Research Council (VR) through Grant No. 2020-00214 and the European Research Council (ERC) under the European Union’s Horizon 2020 research and innovation program (Grant Agreement No.~101001902).
This work was supported by a Simons Investigator Award (Grant No. 511029).
The research of T.K.K. was funded by the Wenner-Gren Foundations.

\appendix

\section{Gradient expansion of the shell index}
\label{app:gradient_expansion}

The shell index is defined as
\begin{equation}
    \mathcal{I}_{\rm shell} = -2\,\mathrm{Tr}\!\left(C\,\rho_\mathcal{A}[\rho_\mathcal{A},\theta]\right),
    \label{eq:appendix_shell_index}
\end{equation}
where $\theta(\mathbf{x})$ is a smooth function and $\rho_\mathcal{A}$ is exponentially localized with a correlation length $\xi$.
 If the characteristic length scale \(\Gamma\) over which \(\theta\) varies is large compared to \(\xi\),
then the commutator \([\rho_\mathcal{A},\theta]\) can be expanded in gradients of \(\theta\).
Let $x=(x_1,\dots,x_d)$ and $y=(y_1,\dots,y_d)$ denote the position eigenvalues in $d$,  spatial dimensions, and adopt the Einstein summation convention over repeated spatial indices. 
The commutator has the matrix elements
\begin{equation}
    \langle x|[\rho_\mathcal{A},\theta]|y\rangle 
    = \rho_\mathcal{A}(x,y)\,\big(\theta(y)-\theta(x)\big).
\end{equation}
If $\theta(x)$ varies smoothly on length scales $\gg \xi$,  then a gradient expansion gives
\begin{align}
    \theta(y)-\theta(x)
    &= (y-x)_i\,\partial_i \theta(x) \nonumber \\
    &\quad + \tfrac{1}{2}(y_i-x_i)(y_j-x_j)\,\partial_i\partial_j \theta(x) \nonumber \\
    &\quad + \mathcal{O}\!\big(|y-x|^3 \,\partial^3\theta\big).
\end{align}
Using
\begin{align}
    \langle x|[\rho_\mathcal{A},X_i]|y\rangle &= (y_i-x_i)\,\rho_\mathcal{A}(x,y), \\
    \langle x|[[\rho_\mathcal{A},X_i],X_j]|y\rangle &= (y_i-x_i)(y_j-x_j)\,\rho_\mathcal{A}(x,y),
\end{align}
the commutator can be written, to second order in gradients, as
\begin{align}
    [\rho_\mathcal{A},\theta]=
    & [\rho_\mathcal{A},X_i]\,\partial_i \theta \nonumber \\
    &+\tfrac{1}{2}\,[[\rho_\mathcal{A},X_i],X_j]\,\partial_i\partial_j \theta \nonumber + \mathcal{O}\!\big(\xi^3 \partial^3 \theta\big).
\end{align}
The matrix elements of $\rho_\mathcal{A}$ decay as $\rho_\mathcal{A}(x,y)\sim e^{-|x-y|/\xi}$, while the $n$th derivative of $\theta$ scales as $\partial^n\theta \sim 1/\Gamma^n$. Consequently, the contribution of the $n$th order term in the expansion is proportional to $(\xi/\Gamma)^{n}$. In the regime $\Gamma \gg \xi$, this constitutes a small parameter, allowing us to keep only the first term at leading order.
Substituting this dominant term into Eq.~\eqref{eq:appendix_shell_index} for the shell index results in
\begin{equation}
    \mathcal{I}_{\rm shell} \;\approx\;
    -2\,\mathrm{Tr}\!\left(C\,\rho_\mathcal{A}[\rho_\mathcal{A},X_i]\right)\,\partial_i\theta.
\end{equation}

\section{The equivalence between the shell index and the chiral local marker in one-dimensional chains with translation invariance}
\label{app:chiral_marker} 
The local chiral marker is a real space topological invariant formulated in terms of the one-particle density matrix. 
It is a $\mathbb{Z}$ invariant characterizing the topology of states in odd spatial dimensions with a chiral constraint.
In one spatial dimension, the local chiral marker is defined as
\begin{equation}
\nu(x)=-2 \sum_{\alpha}\langle \alpha,x|\rho C X\rho|\alpha, x\rangle
\label{eq:appendix_chiral_marker}
\end{equation}
where the trace is over any internal degrees of freedom,  and $\rho$, $C$, and $X$ are the one-particle density matrix, chiral operator, and position operator.
In one dimension, the chiral marker, with respect to the restricted density matrix over a region $\mathcal{A}$ is equal to the shell index in the same region.
In one dimension the chiral marker, with respect to the restricted density matrix over a region $\mathcal{A}$ is equal to the shell index in the same region.
The shell index is, by expanding the commutator $[\rho_{\mathcal{A}}, \theta]$ first order in $\theta(x)-\theta(y)$, expressed as

\begin{equation}
\label{eq:appendix shell}
\mathcal{I}_{\text{shell}} = -2 \, \text{Tr} \left( C \, \rho_{\mathcal{A}} \, [\rho_{\mathcal{A}}, X] \, \partial_x \theta \right),
\end{equation}
where \( \theta \) is diagonal in position space,
\begin{equation}
\theta = \sum_x \theta(x) |x\rangle\langle x|,
\end{equation}
such that the derivative acts as
\begin{equation}
\partial_x \theta = \sum_x \partial_x \theta(x) |x\rangle\langle x|.
\end{equation}
By expressing the full Hilbert space trace Tr as a sum over positions \(x\), and a trace tr over internal degrees of freedom,
\begin{equation}
\text{Tr}(\cdots) = \sum_x \text{tr} \langle x | (\cdots) | x \rangle,
\end{equation}
the shell index becomes
\begin{equation}
\mathcal{I}_{\text{shell}} = -2 \sum_x \partial_x \theta(x) \, \text{tr} \langle x | C \, \rho_{\mathcal{A}} \, [\rho_{\mathcal{A}}, X] | x \rangle,
\label{eq:ap-commutator}
\end{equation}
where the factor \( \partial_x \theta(x) \) is a scalar at each site, allowing it to be pulled outside of the internal trace.
Since $\partial_x\theta(x)$ is nonzero only far from the boundary of $\mathcal A$, the restricted projector $\rho_{\mathcal A}$ may, in this region, be replaced by the bulk projector $\rho$.
Using $\rho^2=\rho$, the first term in the commutator in Eq.~(\ref{eq:ap-commutator}) is therefore proportional to
\begin{equation}
    \text{tr}\braket{x|C\rho_\mathcal{A}\rho_\mathcal{A} X|x}=\text{tr}\braket{x|C\rho X|x}.
    \label{Ap-Eq:trace}
\end{equation}
A constant shift of the position operator in Eq.~\eqref{Ap-Eq:trace}, $X\to X+x_0$, introduces a term $x_0\,\mathrm{tr}\,\langle x|C\rho|x\rangle$.
Since the trace applies over the internal degrees of freedom, it is taken inside the expectation value.
Due to the chiral constraint $\{\rho, C\}=C$,
\begin{equation}
    \text{tr} C\rho= -\text{tr} \rho C + \text{tr} C=- \text{tr}\rho C=-\text{tr}C\rho,
\end{equation}
such that
\begin{equation}
x_0\,\mathrm{tr}\,\langle x|C\rho|x\rangle=0,
\end{equation}
which means that the trace in Eq.~\eqref{Ap-Eq:trace} is invariant under shifts of the origin of $X$.
By choosing the origin at the site $x$, that is, replacing $X$ by $X-x$, and using that $X|x\rangle=x|x\rangle$, one has $(X-x)|x\rangle=0$, and therefore
\begin{equation}
\mathrm{tr}\,\langle x|C\rho X|x\rangle
=
\mathrm{tr}\,\langle x|C\rho (X-x)|x\rangle
=
0.
\end{equation}

The shell index in Eq.\eqref{eq:ap-commutator} therfore reduces to
\begin{equation}
\mathcal{I}_{\text{shell}}
= 2 \sum_x \partial_x \theta(x)\,\mathrm{tr}\,\langle x |  C \rho_{\mathcal A} X \rho_{\mathcal A} | x \rangle .
\end{equation}
By applying the chiral constraint $\{\rho_{\mathcal A},C\}=C$, using $[C,X]=0$, and invoking the translational invariance of the expectation value $\langle x | XC \rho_{\mathcal A} | x \rangle$ far from the boundary of $\mathcal A$, the shell index becomes
\begin{equation}
\mathcal{I}_{\text{shell}}
= -2 \sum_x \partial_x \theta(x)\,\mathrm{tr}\,\langle x |   \rho_{\mathcal A} C X \rho_{\mathcal A} | x \rangle .
\end{equation}
As a result, the shell index is proportional to the local chiral marker defined in Eq.~\eqref{eq:appendix_chiral_marker},
\begin{equation}
\mathcal{I}_{\text{shell}} \approx  \sum_x \partial_x \theta(x)\,\nu(x).
\end{equation}
The local chiral marker is constant throughout the bulk of translation invariant chains, and the derivative $\partial_x \theta(x)$ is only nonzero at the shell boundary, which also lies in the bulk of the chain.
This means that the shell index reduces to  
\begin{equation}
\mathcal{I}_{\text{shell}} = \nu \sum_x \partial_x \theta(x),
\end{equation}
where,  using a discrete integration by parts,
\begin{equation}
\sum_x \partial_x \theta(x) = \theta(x_{\rm{R}}) - \theta(x_{\rm{L}}),
\end{equation}
where $x_{\rm{R}}$, and $x_{\rm{L}}$ are the right and left ends of the chain.
Since \( \theta \) is defined to interpolate between \(0\) and \(1\) across the shell, 
\begin{equation}
\sum_x \partial_x \theta(x) = \pm 1.
\end{equation}
Consequently, up to a sign factor, the shell index is equivalent to the local chiral marker
\begin{equation}
\mathcal{I}_{\text{shell}} = \pm \nu.
\end{equation}

\nocite{zenodo-repo}

\bibliography{refs}

\begin{thebibliography}{97}%
\makeatletter
\providecommand \@ifxundefined [1]{%
 \@ifx{#1\undefined}
}%
\providecommand \@ifnum [1]{%
 \ifnum #1\expandafter \@firstoftwo
 \else \expandafter \@secondoftwo
 \fi
}%
\providecommand \@ifx [1]{%
 \ifx #1\expandafter \@firstoftwo
 \else \expandafter \@secondoftwo
 \fi
}%
\providecommand \natexlab [1]{#1}%
\providecommand \enquote  [1]{``#1''}%
\providecommand \bibnamefont  [1]{#1}%
\providecommand \bibfnamefont [1]{#1}%
\providecommand \citenamefont [1]{#1}%
\providecommand \href@noop [0]{\@secondoftwo}%
\providecommand \href [0]{\begingroup \@sanitize@url \@href}%
\providecommand \@href[1]{\@@startlink{#1}\@@href}%
\providecommand \@@href[1]{\endgroup#1\@@endlink}%
\providecommand \@sanitize@url [0]{\catcode `\\12\catcode `\$12\catcode
  `\&12\catcode `\#12\catcode `\^12\catcode `\_12\catcode `\%12\relax}%
\providecommand \@@startlink[1]{}%
\providecommand \@@endlink[0]{}%
\providecommand \url  [0]{\begingroup\@sanitize@url \@url }%
\providecommand \@url [1]{\endgroup\@href {#1}{\urlprefix }}%
\providecommand \urlprefix  [0]{URL }%
\providecommand \Eprint [0]{\href }%
\providecommand \doibase [0]{https://doi.org/}%
\providecommand \selectlanguage [0]{\@gobble}%
\providecommand \bibinfo  [0]{\@secondoftwo}%
\providecommand \bibfield  [0]{\@secondoftwo}%
\providecommand \translation [1]{[#1]}%
\providecommand \BibitemOpen [0]{}%
\providecommand \bibitemStop [0]{}%
\providecommand \bibitemNoStop [0]{.\EOS\space}%
\providecommand \EOS [0]{\spacefactor3000\relax}%
\providecommand \BibitemShut  [1]{\csname bibitem#1\endcsname}%
\let\auto@bib@innerbib\@empty
\bibitem [{\citenamefont {Schnyder}\ \emph {et~al.}(2008)\citenamefont
  {Schnyder}, \citenamefont {Ryu}, \citenamefont {Furusaki},\ and\
  \citenamefont {Ludwig}}]{schnyder08}%
  \BibitemOpen
  \bibfield  {author} {\bibinfo {author} {\bibfnamefont {A.~P.}\ \bibnamefont
  {Schnyder}}, \bibinfo {author} {\bibfnamefont {S.}~\bibnamefont {Ryu}},
  \bibinfo {author} {\bibfnamefont {A.}~\bibnamefont {Furusaki}},\ and\
  \bibinfo {author} {\bibfnamefont {A.~W.~W.}\ \bibnamefont {Ludwig}},\
  }\bibfield  {title} {\bibinfo {title} {Classification of topological
  insulators and superconductors in three spatial dimensions},\ }\href
  {https://doi.org/10.1103/PhysRevB.78.195125} {\bibfield  {journal} {\bibinfo
  {journal} {Phys. Rev. B}\ }\textbf {\bibinfo {volume} {78}},\ \bibinfo
  {pages} {195125} (\bibinfo {year} {2008})}\BibitemShut {NoStop}%
\bibitem [{\citenamefont {Kitaev}(2009)}]{kitaev09}%
  \BibitemOpen
  \bibfield  {author} {\bibinfo {author} {\bibfnamefont {A.}~\bibnamefont
  {Kitaev}},\ }\bibfield  {title} {\bibinfo {title} {Periodic table for
  topological insulators and superconductors},\ }\href
  {https://doi.org/10.1063/1.3149495} {\bibfield  {journal} {\bibinfo
  {journal} {AIP Conf. Proc.}\ }\textbf {\bibinfo {volume} {1134}},\ \bibinfo
  {pages} {22} (\bibinfo {year} {2009})}\BibitemShut {NoStop}%
\bibitem [{\citenamefont {Ryu}\ \emph {et~al.}(2010)\citenamefont {Ryu},
  \citenamefont {Schnyder}, \citenamefont {Furusaki},\ and\ \citenamefont
  {Ludwig}}]{Ryu2010}%
  \BibitemOpen
  \bibfield  {author} {\bibinfo {author} {\bibfnamefont {S.}~\bibnamefont
  {Ryu}}, \bibinfo {author} {\bibfnamefont {A.~P.}\ \bibnamefont {Schnyder}},
  \bibinfo {author} {\bibfnamefont {A.}~\bibnamefont {Furusaki}},\ and\
  \bibinfo {author} {\bibfnamefont {A.~W.~W.}\ \bibnamefont {Ludwig}},\
  }\bibfield  {title} {\bibinfo {title} {{Topological insulators and
  superconductors: tenfold way and dimensional hierarchy}},\ }\href
  {https://doi.org/10.1088/1367-2630/12/6/065010} {\bibfield  {journal}
  {\bibinfo  {journal} {New J. Phys.}\ }\textbf {\bibinfo {volume} {12}},\
  \bibinfo {pages} {065010} (\bibinfo {year} {2010})}\BibitemShut {NoStop}%
\bibitem [{\citenamefont {Ludwig}(2015)}]{ludwig15}%
  \BibitemOpen
  \bibfield  {author} {\bibinfo {author} {\bibfnamefont {A.~W.~W.}\
  \bibnamefont {Ludwig}},\ }\bibfield  {title} {\bibinfo {title} {Topological
  phases: classification of topological insulators and superconductors of
  non-interacting fermions, and beyond},\ }\href
  {https://doi.org/10.1088/0031-8949/2015/t168/014001} {\bibfield  {journal}
  {\bibinfo  {journal} {Phys. Scr.}\ }\textbf {\bibinfo {volume} {T168}},\
  \bibinfo {pages} {014001} (\bibinfo {year} {2015})}\BibitemShut {NoStop}%
\bibitem [{\citenamefont {Cartan}(1926)}]{cartan26}%
  \BibitemOpen
  \bibfield  {author} {\bibinfo {author} {\bibfnamefont {E.}~\bibnamefont
  {Cartan}},\ }\bibfield  {title} {\bibinfo {title} {Sur une classe remarquable
  d'espaces de {R}iemann},\ }\href {https://doi.org/10.24033/bsmf.1105}
  {\bibfield  {journal} {\bibinfo  {journal} {Bull. Soc. Math. France}\
  }\textbf {\bibinfo {volume} {54}},\ \bibinfo {pages} {214} (\bibinfo {year}
  {1926})}\BibitemShut {NoStop}%
\bibitem [{\citenamefont {Zirnbauer}(1996)}]{zirnbauer96}%
  \BibitemOpen
  \bibfield  {author} {\bibinfo {author} {\bibfnamefont {M.~R.}\ \bibnamefont
  {Zirnbauer}},\ }\bibfield  {title} {\bibinfo {title} {Riemannian symmetric
  superspaces and their origin in random matrix theory},\ }\href
  {https://doi.org/10.1063/1.531675} {\bibfield  {journal} {\bibinfo  {journal}
  {J. Math. Phys.}\ }\textbf {\bibinfo {volume} {37}},\ \bibinfo {pages} {4986}
  (\bibinfo {year} {1996})}\BibitemShut {NoStop}%
\bibitem [{\citenamefont {Altland}\ and\ \citenamefont
  {Zirnbauer}(1997)}]{Altland97}%
  \BibitemOpen
  \bibfield  {author} {\bibinfo {author} {\bibfnamefont {A.}~\bibnamefont
  {Altland}}\ and\ \bibinfo {author} {\bibfnamefont {M.~R.}\ \bibnamefont
  {Zirnbauer}},\ }\bibfield  {title} {\bibinfo {title} {Nonstandard symmetry
  classes in mesoscopic normal-superconducting hybrid structures},\ }\href
  {https://doi.org/10.1103/PhysRevB.55.1142} {\bibfield  {journal} {\bibinfo
  {journal} {Phys. Rev. B}\ }\textbf {\bibinfo {volume} {55}},\ \bibinfo
  {pages} {1142} (\bibinfo {year} {1997})}\BibitemShut {NoStop}%
\bibitem [{\citenamefont {Ando}\ and\ \citenamefont {Fu}(2015)}]{Ando_2015}%
  \BibitemOpen
  \bibfield  {author} {\bibinfo {author} {\bibfnamefont {Y.}~\bibnamefont
  {Ando}}\ and\ \bibinfo {author} {\bibfnamefont {L.}~\bibnamefont {Fu}},\
  }\bibfield  {title} {\bibinfo {title} {Topological crystalline insulators and
  topological superconductors: From concepts to materials},\ }\href
  {https://doi.org/10.1146/annurev-conmatphys-031214-014501} {\bibfield
  {journal} {\bibinfo  {journal} {Annu. Rev. Condens. Matt. Phys.}\ }\textbf
  {\bibinfo {volume} {6}},\ \bibinfo {pages} {361–381} (\bibinfo {year}
  {2015})}\BibitemShut {NoStop}%
\bibitem [{\citenamefont {Neupert}\ and\ \citenamefont
  {Schindler}(2018)}]{Neupert_2018}%
  \BibitemOpen
  \bibfield  {author} {\bibinfo {author} {\bibfnamefont {T.}~\bibnamefont
  {Neupert}}\ and\ \bibinfo {author} {\bibfnamefont {F.}~\bibnamefont
  {Schindler}},\ }\bibinfo {title} {Topological crystalline insulators},\ in\
  \href {https://doi.org/10.1007/978-3-319-76388-0_2} {\emph {\bibinfo
  {booktitle} {Topological Matter}}}\ (\bibinfo  {publisher} {Springer
  International Publishing},\ \bibinfo {year} {2018})\ p.\ \bibinfo {pages}
  {31–61}\BibitemShut {NoStop}%
\bibitem [{\citenamefont {Benalcazar}\ \emph
  {et~al.}(2017{\natexlab{a}})\citenamefont {Benalcazar}, \citenamefont
  {Bernevig},\ and\ \citenamefont {Hughes}}]{Benalcazar2017quantized}%
  \BibitemOpen
  \bibfield  {author} {\bibinfo {author} {\bibfnamefont {W.~A.}\ \bibnamefont
  {Benalcazar}}, \bibinfo {author} {\bibfnamefont {B.~A.}\ \bibnamefont
  {Bernevig}},\ and\ \bibinfo {author} {\bibfnamefont {T.~L.}\ \bibnamefont
  {Hughes}},\ }\bibfield  {title} {\bibinfo {title} {Quantized electric
  multipole insulators},\ }\href {https://doi.org/10.1126/science.aah6442}
  {\bibfield  {journal} {\bibinfo  {journal} {Science}\ }\textbf {\bibinfo
  {volume} {357}},\ \bibinfo {pages} {61} (\bibinfo {year}
  {2017}{\natexlab{a}})}\BibitemShut {NoStop}%
\bibitem [{\citenamefont {Langbehn}\ \emph {et~al.}(2017)\citenamefont
  {Langbehn}, \citenamefont {Peng}, \citenamefont {Trifunovic}, \citenamefont
  {von Oppen},\ and\ \citenamefont
  {Brouwer}}]{langbehn2017reflection-symmetric}%
  \BibitemOpen
  \bibfield  {author} {\bibinfo {author} {\bibfnamefont {J.}~\bibnamefont
  {Langbehn}}, \bibinfo {author} {\bibfnamefont {Y.}~\bibnamefont {Peng}},
  \bibinfo {author} {\bibfnamefont {L.}~\bibnamefont {Trifunovic}}, \bibinfo
  {author} {\bibfnamefont {F.}~\bibnamefont {von Oppen}},\ and\ \bibinfo
  {author} {\bibfnamefont {P.~W.}\ \bibnamefont {Brouwer}},\ }\bibfield
  {title} {\bibinfo {title} {Reflection-symmetric second-order topological
  insulators and superconductors},\ }\href
  {https://doi.org/10.1103/PhysRevLett.119.246401} {\bibfield  {journal}
  {\bibinfo  {journal} {Phys. Rev. Lett.}\ }\textbf {\bibinfo {volume} {119}},\
  \bibinfo {pages} {246401} (\bibinfo {year} {2017})}\BibitemShut {NoStop}%
\bibitem [{\citenamefont {Geier}\ \emph {et~al.}(2018)\citenamefont {Geier},
  \citenamefont {Trifunovic}, \citenamefont {Hoskam},\ and\ \citenamefont
  {Brouwer}}]{Geier2018}%
  \BibitemOpen
  \bibfield  {author} {\bibinfo {author} {\bibfnamefont {M.}~\bibnamefont
  {Geier}}, \bibinfo {author} {\bibfnamefont {L.}~\bibnamefont {Trifunovic}},
  \bibinfo {author} {\bibfnamefont {M.}~\bibnamefont {Hoskam}},\ and\ \bibinfo
  {author} {\bibfnamefont {P.~W.}\ \bibnamefont {Brouwer}},\ }\bibfield
  {title} {\bibinfo {title} {Second-order topological insulators and
  superconductors with an order-two crystalline symmetry},\ }\href
  {https://doi.org/10.1103/PhysRevB.97.205135} {\bibfield  {journal} {\bibinfo
  {journal} {Phys. Rev. B}\ }\textbf {\bibinfo {volume} {97}},\ \bibinfo
  {pages} {205135} (\bibinfo {year} {2018})}\BibitemShut {NoStop}%
\bibitem [{\citenamefont {Schindler}\ \emph {et~al.}(2018)\citenamefont
  {Schindler}, \citenamefont {Cook}, \citenamefont {Vergniory}, \citenamefont
  {Wang}, \citenamefont {Parkin}, \citenamefont {Bernevig},\ and\ \citenamefont
  {Neupert}}]{Schindler2018}%
  \BibitemOpen
  \bibfield  {author} {\bibinfo {author} {\bibfnamefont {F.}~\bibnamefont
  {Schindler}}, \bibinfo {author} {\bibfnamefont {A.~M.}\ \bibnamefont {Cook}},
  \bibinfo {author} {\bibfnamefont {M.~G.}\ \bibnamefont {Vergniory}}, \bibinfo
  {author} {\bibfnamefont {Z.}~\bibnamefont {Wang}}, \bibinfo {author}
  {\bibfnamefont {S.~S.~P.}\ \bibnamefont {Parkin}}, \bibinfo {author}
  {\bibfnamefont {B.~A.}\ \bibnamefont {Bernevig}},\ and\ \bibinfo {author}
  {\bibfnamefont {T.}~\bibnamefont {Neupert}},\ }\bibfield  {title} {\bibinfo
  {title} {Higher-order topological insulators},\ }\href
  {https://www.science.org/doi/10.1126/sciadv.aat0346} {\bibfield  {journal}
  {\bibinfo  {journal} {Sci. Adv.}\ }\textbf {\bibinfo {volume} {4}} (\bibinfo
  {year} {2018})}\BibitemShut {NoStop}%
\bibitem [{\citenamefont {Trifunovic}\ and\ \citenamefont
  {Brouwer}(2019)}]{Trifunovic2019}%
  \BibitemOpen
  \bibfield  {author} {\bibinfo {author} {\bibfnamefont {L.}~\bibnamefont
  {Trifunovic}}\ and\ \bibinfo {author} {\bibfnamefont {P.~W.}\ \bibnamefont
  {Brouwer}},\ }\bibfield  {title} {\bibinfo {title} {Higher-order
  bulk-boundary correspondence for topological crystalline phases},\ }\href
  {https://doi.org/10.1103/PhysRevX.9.011012} {\bibfield  {journal} {\bibinfo
  {journal} {Phys. Rev. X}\ }\textbf {\bibinfo {volume} {9}},\ \bibinfo {pages}
  {011012} (\bibinfo {year} {2019})}\BibitemShut {NoStop}%
\bibitem [{\citenamefont {Trifunovic}\ and\ \citenamefont
  {Brouwer}(2021)}]{Trifunovic2021}%
  \BibitemOpen
  \bibfield  {author} {\bibinfo {author} {\bibfnamefont {L.}~\bibnamefont
  {Trifunovic}}\ and\ \bibinfo {author} {\bibfnamefont {P.~W.}\ \bibnamefont
  {Brouwer}},\ }\bibfield  {title} {\bibinfo {title} {Higher-order topological
  band structures},\ }\href
  {https://onlinelibrary.wiley.com/doi/abs/10.1002/pssb.202000090} {\bibfield
  {journal} {\bibinfo  {journal} {Phys. Status Solidi B}\ }\textbf {\bibinfo
  {volume} {258}} (\bibinfo {year} {2021})}\BibitemShut {NoStop}%
\bibitem [{\citenamefont {Chaou}\ \emph {et~al.}(2023)\citenamefont {Chaou},
  \citenamefont {Brouwer},\ and\ \citenamefont {Sedlmayr}}]{Chaou2023}%
  \BibitemOpen
  \bibfield  {author} {\bibinfo {author} {\bibfnamefont {A.~Y.}\ \bibnamefont
  {Chaou}}, \bibinfo {author} {\bibfnamefont {P.~W.}\ \bibnamefont {Brouwer}},\
  and\ \bibinfo {author} {\bibfnamefont {N.}~\bibnamefont {Sedlmayr}},\
  }\bibfield  {title} {\bibinfo {title} {Hinge states of second-order
  topological insulators as a mach-zehnder interferometer},\ }\href
  {https://doi.org/10.1103/PhysRevB.107.035430} {\bibfield  {journal} {\bibinfo
   {journal} {Phys. Rev. B}\ }\textbf {\bibinfo {volume} {107}},\ \bibinfo
  {pages} {035430} (\bibinfo {year} {2023})}\BibitemShut {NoStop}%
\bibitem [{\citenamefont {Xie}\ \emph {et~al.}(2021)\citenamefont {Xie},
  \citenamefont {Wang}, \citenamefont {Zhang}, \citenamefont {Zhan},
  \citenamefont {Jiang}, \citenamefont {Lu},\ and\ \citenamefont
  {Chen}}]{xie2021higher}%
  \BibitemOpen
  \bibfield  {author} {\bibinfo {author} {\bibfnamefont {B.}~\bibnamefont
  {Xie}}, \bibinfo {author} {\bibfnamefont {H.-X.}\ \bibnamefont {Wang}},
  \bibinfo {author} {\bibfnamefont {X.}~\bibnamefont {Zhang}}, \bibinfo
  {author} {\bibfnamefont {P.}~\bibnamefont {Zhan}}, \bibinfo {author}
  {\bibfnamefont {J.-H.}\ \bibnamefont {Jiang}}, \bibinfo {author}
  {\bibfnamefont {M.}~\bibnamefont {Lu}},\ and\ \bibinfo {author}
  {\bibfnamefont {Y.}~\bibnamefont {Chen}},\ }\bibfield  {title} {\bibinfo
  {title} {Higher-order band topology},\ }\href
  {https://www.nature.com/articles/s41586-021-04002-3} {\bibfield  {journal}
  {\bibinfo  {journal} {Nat. Rev. Phys.}\ }\textbf {\bibinfo {volume} {3}},\
  \bibinfo {pages} {520} (\bibinfo {year} {2021})}\BibitemShut {NoStop}%
\bibitem [{\citenamefont {Varjas}\ \emph {et~al.}(2019)\citenamefont {Varjas},
  \citenamefont {Lau}, \citenamefont {P\"oyh\"onen}, \citenamefont {Akhmerov},
  \citenamefont {Pikulin},\ and\ \citenamefont {Fulga}}]{Varjas2019}%
  \BibitemOpen
  \bibfield  {author} {\bibinfo {author} {\bibfnamefont {D.}~\bibnamefont
  {Varjas}}, \bibinfo {author} {\bibfnamefont {A.}~\bibnamefont {Lau}},
  \bibinfo {author} {\bibfnamefont {K.}~\bibnamefont {P\"oyh\"onen}}, \bibinfo
  {author} {\bibfnamefont {A.~R.}\ \bibnamefont {Akhmerov}}, \bibinfo {author}
  {\bibfnamefont {D.~I.}\ \bibnamefont {Pikulin}},\ and\ \bibinfo {author}
  {\bibfnamefont {I.~C.}\ \bibnamefont {Fulga}},\ }\bibfield  {title} {\bibinfo
  {title} {Topological phases without crystalline counterparts},\ }\href
  {https://doi.org/10.1103/PhysRevLett.123.196401} {\bibfield  {journal}
  {\bibinfo  {journal} {Phys. Rev. Lett.}\ }\textbf {\bibinfo {volume} {123}},\
  \bibinfo {pages} {196401} (\bibinfo {year} {2019})}\BibitemShut {NoStop}%
\bibitem [{\citenamefont {Agarwala}\ \emph {et~al.}(2020)\citenamefont
  {Agarwala}, \citenamefont {Juri\ifmmode \check{c}\else
  \v{c}\fi{}i\ifmmode~\acute{c}\else \'{c}\fi{}},\ and\ \citenamefont
  {Roy}}]{Agarwala2020}%
  \BibitemOpen
  \bibfield  {author} {\bibinfo {author} {\bibfnamefont {A.}~\bibnamefont
  {Agarwala}}, \bibinfo {author} {\bibfnamefont {V.}~\bibnamefont {Juri\ifmmode
  \check{c}\else \v{c}\fi{}i\ifmmode~\acute{c}\else \'{c}\fi{}}},\ and\
  \bibinfo {author} {\bibfnamefont {B.}~\bibnamefont {Roy}},\ }\bibfield
  {title} {\bibinfo {title} {Higher-order topological insulators in amorphous
  solids},\ }\href {https://doi.org/10.1103/PhysRevResearch.2.012067}
  {\bibfield  {journal} {\bibinfo  {journal} {Phys. Rev. Res.}\ }\textbf
  {\bibinfo {volume} {2}},\ \bibinfo {pages} {012067} (\bibinfo {year}
  {2020})}\BibitemShut {NoStop}%
\bibitem [{\citenamefont {Tao}\ \emph {et~al.}(2023)\citenamefont {Tao},
  \citenamefont {Wang},\ and\ \citenamefont {Xu}}]{Tao2023}%
  \BibitemOpen
  \bibfield  {author} {\bibinfo {author} {\bibfnamefont {Y.-L.}\ \bibnamefont
  {Tao}}, \bibinfo {author} {\bibfnamefont {J.-H.}\ \bibnamefont {Wang}},\ and\
  \bibinfo {author} {\bibfnamefont {Y.}~\bibnamefont {Xu}},\ }\bibfield
  {title} {\bibinfo {title} {{Average symmetry protected higher-order
  topological amorphous insulators}},\ }\href
  {https://doi.org/10.21468/SciPostPhys.15.5.193} {\bibfield  {journal}
  {\bibinfo  {journal} {SciPost Phys.}\ }\textbf {\bibinfo {volume} {15}},\
  \bibinfo {pages} {193} (\bibinfo {year} {2023})}\BibitemShut {NoStop}%
\bibitem [{\citenamefont {Peng}\ \emph {et~al.}(2022)\citenamefont {Peng},
  \citenamefont {Hua}, \citenamefont {Chen}, \citenamefont {Liu}, \citenamefont
  {Huang},\ and\ \citenamefont {Zhou}}]{Peng_2022}%
  \BibitemOpen
  \bibfield  {author} {\bibinfo {author} {\bibfnamefont {T.}~\bibnamefont
  {Peng}}, \bibinfo {author} {\bibfnamefont {C.-B.}\ \bibnamefont {Hua}},
  \bibinfo {author} {\bibfnamefont {R.}~\bibnamefont {Chen}}, \bibinfo {author}
  {\bibfnamefont {Z.-R.}\ \bibnamefont {Liu}}, \bibinfo {author} {\bibfnamefont
  {H.-M.}\ \bibnamefont {Huang}},\ and\ \bibinfo {author} {\bibfnamefont
  {B.}~\bibnamefont {Zhou}},\ }\bibfield  {title} {\bibinfo {title}
  {Density-driven higher-order topological phase transitions in amorphous
  solids},\ }\href {https://doi.org/10.1103/PhysRevB.106.125310} {\bibfield
  {journal} {\bibinfo  {journal} {Phys. Rev. B}\ }\textbf {\bibinfo {volume}
  {106}},\ \bibinfo {pages} {125310} (\bibinfo {year} {2022})}\BibitemShut
  {NoStop}%
\bibitem [{\citenamefont {Chaou}\ \emph {et~al.}(2025)\citenamefont {Chaou},
  \citenamefont {Moreno-Gonzalez}, \citenamefont {Altland},\ and\ \citenamefont
  {Brouwer}}]{Chaou2025}%
  \BibitemOpen
  \bibfield  {author} {\bibinfo {author} {\bibfnamefont {A.~Y.}\ \bibnamefont
  {Chaou}}, \bibinfo {author} {\bibfnamefont {M.}~\bibnamefont
  {Moreno-Gonzalez}}, \bibinfo {author} {\bibfnamefont {A.}~\bibnamefont
  {Altland}},\ and\ \bibinfo {author} {\bibfnamefont {P.~W.}\ \bibnamefont
  {Brouwer}},\ }\bibfield  {title} {\bibinfo {title} {Disordered topological
  crystalline phases},\ }\href {https://doi.org/10.1103/fpc3-bsz6} {\bibfield
  {journal} {\bibinfo  {journal} {Phys. Rev. B}\ }\textbf {\bibinfo {volume}
  {112}},\ \bibinfo {pages} {035167} (\bibinfo {year} {2025})}\BibitemShut
  {NoStop}%
\bibitem [{\citenamefont {Prodan}(2010)}]{Prodan2010}%
  \BibitemOpen
  \bibfield  {author} {\bibinfo {author} {\bibfnamefont {E.}~\bibnamefont
  {Prodan}},\ }\bibfield  {title} {\bibinfo {title} {Non-commutative tools for
  topological insulators},\ }\href
  {https://doi.org/10.1088/1367-2630/12/6/065003} {\bibfield  {journal}
  {\bibinfo  {journal} {New J. Phys.}\ }\textbf {\bibinfo {volume} {12}},\
  \bibinfo {pages} {065003} (\bibinfo {year} {2010})}\BibitemShut {NoStop}%
\bibitem [{\citenamefont {Loring}\ and\ \citenamefont
  {Hastings}(2011)}]{Loring2010}%
  \BibitemOpen
  \bibfield  {author} {\bibinfo {author} {\bibfnamefont {T.~A.}\ \bibnamefont
  {Loring}}\ and\ \bibinfo {author} {\bibfnamefont {M.~B.}\ \bibnamefont
  {Hastings}},\ }\bibfield  {title} {\bibinfo {title} {Disordered topological
  insulators via c*-algebras},\ }\href
  {https://doi.org/10.1209/0295-5075/92/67004} {\bibfield  {journal} {\bibinfo
  {journal} {EPL}\ }\textbf {\bibinfo {volume} {92}},\ \bibinfo {pages} {67004}
  (\bibinfo {year} {2011})}\BibitemShut {NoStop}%
\bibitem [{\citenamefont {Prodan}(2011)}]{Prodan2011}%
  \BibitemOpen
  \bibfield  {author} {\bibinfo {author} {\bibfnamefont {E.}~\bibnamefont
  {Prodan}},\ }\bibfield  {title} {\bibinfo {title} {Disordered topological
  insulators: a non-commutative geometry perspective},\ }\href
  {https://doi.org/10.1088/1751-8113/44/11/113001} {\bibfield  {journal}
  {\bibinfo  {journal} {J. Phys. A}\ }\textbf {\bibinfo {volume} {44}},\
  \bibinfo {pages} {113001} (\bibinfo {year} {2011})}\BibitemShut {NoStop}%
\bibitem [{\citenamefont {Bianco}\ and\ \citenamefont
  {Resta}(2011)}]{Bianco2011}%
  \BibitemOpen
  \bibfield  {author} {\bibinfo {author} {\bibfnamefont {R.}~\bibnamefont
  {Bianco}}\ and\ \bibinfo {author} {\bibfnamefont {R.}~\bibnamefont {Resta}},\
  }\bibfield  {title} {\bibinfo {title} {Mapping topological order in
  coordinate space},\ }\href {https://doi.org/10.1103/PhysRevB.84.241106}
  {\bibfield  {journal} {\bibinfo  {journal} {Phys. Rev. B}\ }\textbf {\bibinfo
  {volume} {84}},\ \bibinfo {pages} {241106} (\bibinfo {year}
  {2011})}\BibitemShut {NoStop}%
\bibitem [{\citenamefont {Loring}(2015)}]{Loring2015}%
  \BibitemOpen
  \bibfield  {author} {\bibinfo {author} {\bibfnamefont {T.~A.}\ \bibnamefont
  {Loring}},\ }\bibfield  {title} {\bibinfo {title} {K-theory and pseudospectra
  for topological insulators},\ }\href
  {https://doi.org/https://doi.org/10.1016/j.aop.2015.02.031} {\bibfield
  {journal} {\bibinfo  {journal} {Ann. Phys.}\ }\textbf {\bibinfo {volume}
  {356}},\ \bibinfo {pages} {383} (\bibinfo {year} {2015})}\BibitemShut
  {NoStop}%
\bibitem [{\citenamefont {Huang}\ and\ \citenamefont {Liu}(2018)}]{Huang2018}%
  \BibitemOpen
  \bibfield  {author} {\bibinfo {author} {\bibfnamefont {H.}~\bibnamefont
  {Huang}}\ and\ \bibinfo {author} {\bibfnamefont {F.}~\bibnamefont {Liu}},\
  }\bibfield  {title} {\bibinfo {title} {Theory of spin bott index for quantum
  spin hall states in nonperiodic systems},\ }\href
  {https://doi.org/10.1103/PhysRevB.98.125130} {\bibfield  {journal} {\bibinfo
  {journal} {Phys. Rev. B}\ }\textbf {\bibinfo {volume} {98}},\ \bibinfo
  {pages} {125130} (\bibinfo {year} {2018})}\BibitemShut {NoStop}%
\bibitem [{\citenamefont {Irsigler}\ \emph {et~al.}(2019)\citenamefont
  {Irsigler}, \citenamefont {Zheng},\ and\ \citenamefont
  {Hofstetter}}]{Hofstetter2019}%
  \BibitemOpen
  \bibfield  {author} {\bibinfo {author} {\bibfnamefont {B.}~\bibnamefont
  {Irsigler}}, \bibinfo {author} {\bibfnamefont {J.-H.}\ \bibnamefont
  {Zheng}},\ and\ \bibinfo {author} {\bibfnamefont {W.}~\bibnamefont
  {Hofstetter}},\ }\bibfield  {title} {\bibinfo {title} {Microscopic
  characteristics and tomography scheme of the local chern marker},\ }\href
  {https://doi.org/10.1103/PhysRevA.100.023610} {\bibfield  {journal} {\bibinfo
   {journal} {Phys. Rev. A}\ }\textbf {\bibinfo {volume} {100}},\ \bibinfo
  {pages} {023610} (\bibinfo {year} {2019})}\BibitemShut {NoStop}%
\bibitem [{\citenamefont {Loring}()}]{Loring2019}%
  \BibitemOpen
  \bibfield  {author} {\bibinfo {author} {\bibfnamefont {T.~A.}\ \bibnamefont
  {Loring}},\ }\href {https://doi.org/10.48550/arXiv.1907.11791} {}\Eprint
  {https://arxiv.org/abs/1907.11791} {arXiv:1907.11791} \BibitemShut {NoStop}%
\bibitem [{\citenamefont {Mondragon-Shem}\ and\ \citenamefont
  {Hughes}(2024)}]{Hughes2019}%
  \BibitemOpen
  \bibfield  {author} {\bibinfo {author} {\bibfnamefont {I.}~\bibnamefont
  {Mondragon-Shem}}\ and\ \bibinfo {author} {\bibfnamefont {T.~L.}\
  \bibnamefont {Hughes}},\ }\bibfield  {title} {\bibinfo {title} {Robust
  topological invariants of topological crystalline phases in the presence of
  impurities},\ }\href {https://doi.org/10.1103/PhysRevB.110.035146} {\bibfield
   {journal} {\bibinfo  {journal} {Phys. Rev. B}\ }\textbf {\bibinfo {volume}
  {110}},\ \bibinfo {pages} {035146} (\bibinfo {year} {2024})}\BibitemShut
  {NoStop}%
\bibitem [{\citenamefont {Jezequel}\ \emph {et~al.}(2022)\citenamefont
  {Jezequel}, \citenamefont {Tauber},\ and\ \citenamefont
  {Delplace}}]{jezequel2022}%
  \BibitemOpen
  \bibfield  {author} {\bibinfo {author} {\bibfnamefont {L.}~\bibnamefont
  {Jezequel}}, \bibinfo {author} {\bibfnamefont {C.}~\bibnamefont {Tauber}},\
  and\ \bibinfo {author} {\bibfnamefont {P.}~\bibnamefont {Delplace}},\
  }\bibfield  {title} {\bibinfo {title} {Estimating bulk and edge topological
  indices in finite open chiral chains},\ }\href
  {https://doi.org/https://doi.org/10.1063/5.0096720} {\bibfield  {journal}
  {\bibinfo  {journal} {J. Math. Phys.}\ }\textbf {\bibinfo {volume} {63}},\
  \bibinfo {pages} {121901} (\bibinfo {year} {2022})}\BibitemShut {NoStop}%
\bibitem [{\citenamefont {Hannukainen}\ \emph {et~al.}(2022)\citenamefont
  {Hannukainen}, \citenamefont {Mart\'{\i}nez}, \citenamefont {Bardarson},\
  and\ \citenamefont {Kvorning}}]{Hannukainen2022}%
  \BibitemOpen
  \bibfield  {author} {\bibinfo {author} {\bibfnamefont {J.~D.}\ \bibnamefont
  {Hannukainen}}, \bibinfo {author} {\bibfnamefont {M.~F.}\ \bibnamefont
  {Mart\'{\i}nez}}, \bibinfo {author} {\bibfnamefont {J.~H.}\ \bibnamefont
  {Bardarson}},\ and\ \bibinfo {author} {\bibfnamefont {T.~K.}\ \bibnamefont
  {Kvorning}},\ }\bibfield  {title} {\bibinfo {title} {Local topological
  markers in odd spatial dimensions and their application to amorphous
  topological matter},\ }\href {https://doi.org/10.1103/PhysRevLett.129.277601}
  {\bibfield  {journal} {\bibinfo  {journal} {Phys. Rev. Lett.}\ }\textbf
  {\bibinfo {volume} {129}},\ \bibinfo {pages} {277601} (\bibinfo {year}
  {2022})}\BibitemShut {NoStop}%
\bibitem [{\citenamefont {Hannukainen}\ \emph {et~al.}(2024)\citenamefont
  {Hannukainen}, \citenamefont {Mart\'{\i}nez}, \citenamefont {Bardarson},\
  and\ \citenamefont {Kvorning}}]{Hannukainen2024}%
  \BibitemOpen
  \bibfield  {author} {\bibinfo {author} {\bibfnamefont {J.~D.}\ \bibnamefont
  {Hannukainen}}, \bibinfo {author} {\bibfnamefont {M.~F.}\ \bibnamefont
  {Mart\'{\i}nez}}, \bibinfo {author} {\bibfnamefont {J.~H.}\ \bibnamefont
  {Bardarson}},\ and\ \bibinfo {author} {\bibfnamefont {T.~K.}\ \bibnamefont
  {Kvorning}},\ }\bibfield  {title} {\bibinfo {title} {Interacting local
  topological markers: A one-particle density matrix approach for
  characterizing the topology of interacting and disordered states},\ }\href
  {https://doi.org/10.1103/PhysRevResearch.6.L032045} {\bibfield  {journal}
  {\bibinfo  {journal} {Phys. Rev. Res.}\ }\textbf {\bibinfo {volume} {6}},\
  \bibinfo {pages} {L032045} (\bibinfo {year} {2024})}\BibitemShut {NoStop}%
\bibitem [{\citenamefont {Marrazzo}\ and\ \citenamefont
  {Resta}(2017)}]{LocalityAHC}%
  \BibitemOpen
  \bibfield  {author} {\bibinfo {author} {\bibfnamefont {A.}~\bibnamefont
  {Marrazzo}}\ and\ \bibinfo {author} {\bibfnamefont {R.}~\bibnamefont
  {Resta}},\ }\bibfield  {title} {\bibinfo {title} {Locality of the anomalous
  hall conductivity},\ }\href {https://doi.org/10.1103/PhysRevB.95.121114}
  {\bibfield  {journal} {\bibinfo  {journal} {Phys. Rev. B}\ }\textbf {\bibinfo
  {volume} {95}},\ \bibinfo {pages} {121114} (\bibinfo {year}
  {2017})}\BibitemShut {NoStop}%
\bibitem [{\citenamefont {Ba\`u}\ and\ \citenamefont
  {Marrazzo}(2024)}]{LocalMarker}%
  \BibitemOpen
  \bibfield  {author} {\bibinfo {author} {\bibfnamefont {N.}~\bibnamefont
  {Ba\`u}}\ and\ \bibinfo {author} {\bibfnamefont {A.}~\bibnamefont
  {Marrazzo}},\ }\bibfield  {title} {\bibinfo {title} {Local chern marker for
  periodic systems},\ }\href {https://doi.org/10.1103/PhysRevB.109.014206}
  {\bibfield  {journal} {\bibinfo  {journal} {Phys. Rev. B}\ }\textbf {\bibinfo
  {volume} {109}},\ \bibinfo {pages} {014206} (\bibinfo {year}
  {2024})}\BibitemShut {NoStop}%
\bibitem [{\citenamefont {Caio}\ \emph {et~al.}(2019)\citenamefont {Caio},
  \citenamefont {Möller}, \citenamefont {Cooper},\ and\ \citenamefont
  {Bhaseen}}]{caio2019topological}%
  \BibitemOpen
  \bibfield  {author} {\bibinfo {author} {\bibfnamefont {M.~D.}\ \bibnamefont
  {Caio}}, \bibinfo {author} {\bibfnamefont {G.}~\bibnamefont {Möller}},
  \bibinfo {author} {\bibfnamefont {N.~R.}\ \bibnamefont {Cooper}},\ and\
  \bibinfo {author} {\bibfnamefont {M.~J.}\ \bibnamefont {Bhaseen}},\
  }\bibfield  {title} {\bibinfo {title} {Topological marker currents in chern
  insulators},\ }\href {https://doi.org/10.1038/s41567-018-0390-7} {\bibfield
  {journal} {\bibinfo  {journal} {Nat. Phys.}\ }\textbf {\bibinfo {volume}
  {15}},\ \bibinfo {pages} {257–261} (\bibinfo {year} {2019})}\BibitemShut
  {NoStop}%
\bibitem [{\citenamefont {d'Ornellas}\ \emph {et~al.}(2022)\citenamefont
  {d'Ornellas}, \citenamefont {Barnett},\ and\ \citenamefont
  {Lee}}]{dOrnellas2022}%
  \BibitemOpen
  \bibfield  {author} {\bibinfo {author} {\bibfnamefont {P.}~\bibnamefont
  {d'Ornellas}}, \bibinfo {author} {\bibfnamefont {R.}~\bibnamefont
  {Barnett}},\ and\ \bibinfo {author} {\bibfnamefont {D.~K.~K.}\ \bibnamefont
  {Lee}},\ }\bibfield  {title} {\bibinfo {title} {Quantized bulk conductivity
  as a local chern marker},\ }\href
  {https://doi.org/10.1103/PhysRevB.106.155124} {\bibfield  {journal} {\bibinfo
   {journal} {Phys. Rev. B}\ }\textbf {\bibinfo {volume} {106}},\ \bibinfo
  {pages} {155124} (\bibinfo {year} {2022})}\BibitemShut {NoStop}%
\bibitem [{\citenamefont {Kitaev}(2006)}]{kitaev2006anyons}%
  \BibitemOpen
  \bibfield  {author} {\bibinfo {author} {\bibfnamefont {A.}~\bibnamefont
  {Kitaev}},\ }\bibfield  {title} {\bibinfo {title} {Anyons in an exactly
  solved model and beyond},\ }\href {https://doi.org/10.1016/j.aop.2005.10.005}
  {\bibfield  {journal} {\bibinfo  {journal} {Ann. Phys.}\ }\textbf {\bibinfo
  {volume} {321}},\ \bibinfo {pages} {2–111} (\bibinfo {year}
  {2006})}\BibitemShut {NoStop}%
\bibitem [{\citenamefont {Loring}\ and\ \citenamefont
  {Schulz-Baldes}(2019)}]{loring2019spectral}%
  \BibitemOpen
  \bibfield  {author} {\bibinfo {author} {\bibfnamefont {T.~A.}\ \bibnamefont
  {Loring}}\ and\ \bibinfo {author} {\bibfnamefont {H.}~\bibnamefont
  {Schulz-Baldes}},\ }\bibfield  {title} {\bibinfo {title} {Spectral flow
  argument localizing an odd index pairing},\ }\href
  {https://doi.org/https://doi.org/10.4153/CMB-2018-013-x} {\bibfield
  {journal} {\bibinfo  {journal} {Canad. Math. Bull.}\ }\textbf {\bibinfo
  {volume} {62}},\ \bibinfo {pages} {373} (\bibinfo {year} {2019})}\BibitemShut
  {NoStop}%
\bibitem [{\citenamefont {Loring}\ and\ \citenamefont
  {Schulz-Baldes}(2020)}]{loring2020spectral}%
  \BibitemOpen
  \bibfield  {author} {\bibinfo {author} {\bibfnamefont {T.~A.}\ \bibnamefont
  {Loring}}\ and\ \bibinfo {author} {\bibfnamefont {H.}~\bibnamefont
  {Schulz-Baldes}},\ }\bibfield  {title} {\bibinfo {title} {The spectral
  localizer for even index pairings},\ }\href
  {https://doi.org/https://doi.org/10.4171/jncg/357} {\bibfield  {journal}
  {\bibinfo  {journal} {J. Noncommuttal. Geom.}\ }\textbf {\bibinfo {volume}
  {14}},\ \bibinfo {pages} {1} (\bibinfo {year} {2020})}\BibitemShut {NoStop}%
\bibitem [{\citenamefont {Schulz-Baldes}\ and\ \citenamefont
  {Stoiber}(2021{\natexlab{a}})}]{schulz2021spectral}%
  \BibitemOpen
  \bibfield  {author} {\bibinfo {author} {\bibfnamefont {H.}~\bibnamefont
  {Schulz-Baldes}}\ and\ \bibinfo {author} {\bibfnamefont {T.}~\bibnamefont
  {Stoiber}},\ }\bibfield  {title} {\bibinfo {title} {The spectral localizer
  for semifinite spectral triples},\ }\href
  {https://doi.org/https://doi.org/10.1090/proc/15230} {\bibfield  {journal}
  {\bibinfo  {journal} {Proc. Am. Math. Soc}\ }\textbf {\bibinfo {volume}
  {149}},\ \bibinfo {pages} {121} (\bibinfo {year}
  {2021}{\natexlab{a}})}\BibitemShut {NoStop}%
\bibitem [{\citenamefont {Schulz-Baldes}\ and\ \citenamefont
  {Stoiber}(2021{\natexlab{b}})}]{schulz2022invariants}%
  \BibitemOpen
  \bibfield  {author} {\bibinfo {author} {\bibfnamefont {H.}~\bibnamefont
  {Schulz-Baldes}}\ and\ \bibinfo {author} {\bibfnamefont {T.}~\bibnamefont
  {Stoiber}},\ }\bibfield  {title} {\bibinfo {title} {Invariants of disordered
  semimetals via the spectral localizer},\ }\href
  {https://doi.org/10.1209/0295-5075/ac1b65} {\bibfield  {journal} {\bibinfo
  {journal} {Europhys. Lett.}\ }\textbf {\bibinfo {volume} {136}},\ \bibinfo
  {pages} {27001} (\bibinfo {year} {2021}{\natexlab{b}})}\BibitemShut {NoStop}%
\bibitem [{\citenamefont {Schulz-Baldes}\ and\ \citenamefont
  {Stoiber}(2023)}]{schulz2023spectral}%
  \BibitemOpen
  \bibfield  {author} {\bibinfo {author} {\bibfnamefont {H.}~\bibnamefont
  {Schulz-Baldes}}\ and\ \bibinfo {author} {\bibfnamefont {T.}~\bibnamefont
  {Stoiber}},\ }\bibfield  {title} {\bibinfo {title} {Spectral localization for
  semimetals and callias operators},\ }\href
  {https://doi.org/https://doi.org/10.1063/5.0093983} {\bibfield  {journal}
  {\bibinfo  {journal} {J. Math. Phys.}\ }\textbf {\bibinfo {volume} {64}},\
  \bibinfo {pages} {8} (\bibinfo {year} {2023})}\BibitemShut {NoStop}%
\bibitem [{\citenamefont {Franca}\ and\ \citenamefont
  {Grushin}(2024)}]{franca2024topological}%
  \BibitemOpen
  \bibfield  {author} {\bibinfo {author} {\bibfnamefont {S.}~\bibnamefont
  {Franca}}\ and\ \bibinfo {author} {\bibfnamefont {A.~G.}\ \bibnamefont
  {Grushin}},\ }\bibfield  {title} {\bibinfo {title} {Topological zero-modes of
  the spectral localizer of trivial metals},\ }\href
  {https://doi.org/10.1103/PhysRevB.109.195107} {\bibfield  {journal} {\bibinfo
   {journal} {Phys. Rev. B}\ }\textbf {\bibinfo {volume} {109}},\ \bibinfo
  {pages} {195107} (\bibinfo {year} {2024})}\BibitemShut {NoStop}%
\bibitem [{\citenamefont {Stoiber}(2025)}]{stoiber2024spectral}%
  \BibitemOpen
  \bibfield  {author} {\bibinfo {author} {\bibfnamefont {T.}~\bibnamefont
  {Stoiber}},\ }\bibfield  {title} {\bibinfo {title} {A spectral localizer
  approach to strong topological invariants in the mobility gap regime},\
  }\href {https://doi.org/https://doi.org/10.1007/s00220-025-05359-6}
  {\bibfield  {journal} {\bibinfo  {journal} {Commun. Math. Phys.}\ }\textbf
  {\bibinfo {volume} {406}},\ \bibinfo {pages} {184} (\bibinfo {year}
  {2025})}\BibitemShut {NoStop}%
\bibitem [{\citenamefont {Doll}\ and\ \citenamefont
  {Schulz-Baldes}(2021)}]{doll2021skew}%
  \BibitemOpen
  \bibfield  {author} {\bibinfo {author} {\bibfnamefont {N.}~\bibnamefont
  {Doll}}\ and\ \bibinfo {author} {\bibfnamefont {H.}~\bibnamefont
  {Schulz-Baldes}},\ }\bibfield  {title} {\bibinfo {title} {Skew localizer and
  z2-flows for real index pairings},\ }\href
  {https://doi.org/https://doi.org/10.1016/j.aim.2021.108038} {\bibfield
  {journal} {\bibinfo  {journal} {Adv. Math.}\ }\textbf {\bibinfo {volume}
  {392}},\ \bibinfo {pages} {108038} (\bibinfo {year} {2021})}\BibitemShut
  {NoStop}%
\bibitem [{\citenamefont {Cerjan}\ \emph {et~al.}(2023)\citenamefont {Cerjan},
  \citenamefont {Koekenbier},\ and\ \citenamefont
  {Schulz-Baldes}}]{cerjan2023spectral}%
  \BibitemOpen
  \bibfield  {author} {\bibinfo {author} {\bibfnamefont {A.}~\bibnamefont
  {Cerjan}}, \bibinfo {author} {\bibfnamefont {L.}~\bibnamefont {Koekenbier}},\
  and\ \bibinfo {author} {\bibfnamefont {H.}~\bibnamefont {Schulz-Baldes}},\
  }\bibfield  {title} {\bibinfo {title} {Spectral localizer for line-gapped
  non-hermitian systems},\ }\href {http://dx.doi.org/10.1063/5.0150995}
  {\bibfield  {journal} {\bibinfo  {journal} {J. Math. Phys.}\ }\textbf
  {\bibinfo {volume} {64}} (\bibinfo {year} {2023})}\BibitemShut {NoStop}%
\bibitem [{\citenamefont {Doll}\ \emph {et~al.}(2025)\citenamefont {Doll},
  \citenamefont {Loring},\ and\ \citenamefont {Schulz-Baldes}}]{doll2024local}%
  \BibitemOpen
  \bibfield  {author} {\bibinfo {author} {\bibfnamefont {N.}~\bibnamefont
  {Doll}}, \bibinfo {author} {\bibfnamefont {T.}~\bibnamefont {Loring}},\ and\
  \bibinfo {author} {\bibfnamefont {H.}~\bibnamefont {Schulz-Baldes}},\
  }\bibfield  {title} {\bibinfo {title} {Topological indices for periodic
  gapped hamiltonians and fuzzy tori},\ }\href
  {https://doi.org/https://doi.org/10.1007/s11040-025-09508-0} {\bibfield
  {journal} {\bibinfo  {journal} {J. Math. Phys. Anal. Geom.}\ }\textbf
  {\bibinfo {volume} {28}},\ \bibinfo {pages} {13} (\bibinfo {year}
  {2025})}\BibitemShut {NoStop}%
\bibitem [{\citenamefont {Schulz-Baldes}(2025)}]{schulz2024topological}%
  \BibitemOpen
  \bibfield  {author} {\bibinfo {author} {\bibfnamefont {H.}~\bibnamefont
  {Schulz-Baldes}},\ }\bibfield  {title} {\bibinfo {title} {Topological indices
  in condensed matter},\ }in\ \href
  {https://doi.org/10.1016/B978-0-323-95703-8.00041-0} {\emph {\bibinfo
  {booktitle} {Encyclopedia of Mathematical Physics}}},\ \bibinfo {editor}
  {edited by\ \bibinfo {editor} {\bibfnamefont {R.}~\bibnamefont {Szabo}}\ and\
  \bibinfo {editor} {\bibfnamefont {M.}~\bibnamefont {Bojowald}}}\ (\bibinfo
  {publisher} {Academic Press},\ \bibinfo {address} {Oxford},\ \bibinfo {year}
  {2025})\ \bibinfo {edition} {second edition}\ ed.,\ pp.\ \bibinfo {pages}
  {17--26}\BibitemShut {NoStop}%
\bibitem [{\citenamefont {Jezequel}\ \emph {et~al.}(2026)\citenamefont
  {Jezequel}, \citenamefont {Bardarson},\ and\ \citenamefont
  {Grushin}}]{jezequel2025localizer}%
  \BibitemOpen
  \bibfield  {author} {\bibinfo {author} {\bibfnamefont {L.}~\bibnamefont
  {Jezequel}}, \bibinfo {author} {\bibfnamefont {J.~H.}\ \bibnamefont
  {Bardarson}},\ and\ \bibinfo {author} {\bibfnamefont {A.~G.}\ \bibnamefont
  {Grushin}},\ }\bibfield  {title} {\bibinfo {title} {{Explicit equivalence
  between the spectral localizer and local Chern and winding markers}},\ }\href
  {https://doi.org/10.21468/SciPostPhys.20.4.118} {\bibfield  {journal}
  {\bibinfo  {journal} {SciPost Phys.}\ }\textbf {\bibinfo {volume} {20}},\
  \bibinfo {pages} {118} (\bibinfo {year} {2026})}\BibitemShut {NoStop}%
\bibitem [{\citenamefont {Penrose}\ and\ \citenamefont
  {Onsager}(1956)}]{Penrose1956}%
  \BibitemOpen
  \bibfield  {author} {\bibinfo {author} {\bibfnamefont {O.}~\bibnamefont
  {Penrose}}\ and\ \bibinfo {author} {\bibfnamefont {L.}~\bibnamefont
  {Onsager}},\ }\bibfield  {title} {\bibinfo {title} {Bose-einstein
  condensation and liquid helium},\ }\href
  {https://doi.org/10.1103/PhysRev.104.576} {\bibfield  {journal} {\bibinfo
  {journal} {Phys. Rev.}\ }\textbf {\bibinfo {volume} {104}},\ \bibinfo {pages}
  {576} (\bibinfo {year} {1956})}\BibitemShut {NoStop}%
\bibitem [{\citenamefont {Koch}\ and\ \citenamefont
  {Goedecker}(2001)}]{Koch_2001}%
  \BibitemOpen
  \bibfield  {author} {\bibinfo {author} {\bibfnamefont {E.}~\bibnamefont
  {Koch}}\ and\ \bibinfo {author} {\bibfnamefont {S.}~\bibnamefont
  {Goedecker}},\ }\bibfield  {title} {\bibinfo {title} {Locality properties and
  wannier functions for interacting systems},\ }\href
  {https://doi.org/10.1016/s0038-1098(01)00192-2} {\bibfield  {journal}
  {\bibinfo  {journal} {Solid State Commun.}\ }\textbf {\bibinfo {volume}
  {119}},\ \bibinfo {pages} {105–109} (\bibinfo {year} {2001})}\BibitemShut
  {NoStop}%
\bibitem [{\citenamefont {Bera}\ \emph {et~al.}(2015)\citenamefont {Bera},
  \citenamefont {Schomerus}, \citenamefont {Heidrich-Meisner},\ and\
  \citenamefont {Bardarson}}]{Bera2015}%
  \BibitemOpen
  \bibfield  {author} {\bibinfo {author} {\bibfnamefont {S.}~\bibnamefont
  {Bera}}, \bibinfo {author} {\bibfnamefont {H.}~\bibnamefont {Schomerus}},
  \bibinfo {author} {\bibfnamefont {F.}~\bibnamefont {Heidrich-Meisner}},\ and\
  \bibinfo {author} {\bibfnamefont {J.~H.}\ \bibnamefont {Bardarson}},\
  }\bibfield  {title} {\bibinfo {title} {Many-body localization characterized
  from a one-particle perspective},\ }\href
  {https://doi.org/10.1103/PhysRevLett.115.046603} {\bibfield  {journal}
  {\bibinfo  {journal} {Phys. Rev. Lett.}\ }\textbf {\bibinfo {volume} {115}},\
  \bibinfo {pages} {046603} (\bibinfo {year} {2015})}\BibitemShut {NoStop}%
\bibitem [{\citenamefont {Bera}\ \emph {et~al.}(2017)\citenamefont {Bera},
  \citenamefont {Martynec}, \citenamefont {Schomerus}, \citenamefont
  {Heidrich-Meisner},\ and\ \citenamefont {Bardarson}}]{bera2017one}%
  \BibitemOpen
  \bibfield  {author} {\bibinfo {author} {\bibfnamefont {S.}~\bibnamefont
  {Bera}}, \bibinfo {author} {\bibfnamefont {T.}~\bibnamefont {Martynec}},
  \bibinfo {author} {\bibfnamefont {H.}~\bibnamefont {Schomerus}}, \bibinfo
  {author} {\bibfnamefont {F.}~\bibnamefont {Heidrich-Meisner}},\ and\ \bibinfo
  {author} {\bibfnamefont {J.~H.}\ \bibnamefont {Bardarson}},\ }\bibfield
  {title} {\bibinfo {title} {One-particle density matrix characterization of
  many-body localization},\ }\href
  {https://onlinelibrary.wiley.com/doi/abs/10.1002/andp.201600356} {\bibfield
  {journal} {\bibinfo  {journal} {Ann. Phys. (Berlin)}\ }\textbf {\bibinfo
  {volume} {529}},\ \bibinfo {pages} {1600356} (\bibinfo {year}
  {2017})}\BibitemShut {NoStop}%
\bibitem [{\citenamefont {Lezama}\ \emph {et~al.}(2017)\citenamefont {Lezama},
  \citenamefont {Bera}, \citenamefont {Schomerus}, \citenamefont
  {Heidrich-Meisner},\ and\ \citenamefont {Bardarson}}]{Lezama2017}%
  \BibitemOpen
  \bibfield  {author} {\bibinfo {author} {\bibfnamefont {T.~L.~M.}\
  \bibnamefont {Lezama}}, \bibinfo {author} {\bibfnamefont {S.}~\bibnamefont
  {Bera}}, \bibinfo {author} {\bibfnamefont {H.}~\bibnamefont {Schomerus}},
  \bibinfo {author} {\bibfnamefont {F.}~\bibnamefont {Heidrich-Meisner}},\ and\
  \bibinfo {author} {\bibfnamefont {J.~H.}\ \bibnamefont {Bardarson}},\
  }\bibfield  {title} {\bibinfo {title} {One-particle density matrix occupation
  spectrum of many-body localized states after a global quench},\ }\href
  {https://doi.org/10.1103/physrevb.96.060202} {\bibfield  {journal} {\bibinfo
  {journal} {Phys. Rev. B}\ }\textbf {\bibinfo {volume} {96}},\ \bibinfo
  {pages} {060202(R)} (\bibinfo {year} {2017})}\BibitemShut {NoStop}%
\bibitem [{\citenamefont {Kells}\ \emph {et~al.}(2018)\citenamefont {Kells},
  \citenamefont {Moran},\ and\ \citenamefont {Meidan}}]{Kells2018}%
  \BibitemOpen
  \bibfield  {author} {\bibinfo {author} {\bibfnamefont {G.}~\bibnamefont
  {Kells}}, \bibinfo {author} {\bibfnamefont {N.}~\bibnamefont {Moran}},\ and\
  \bibinfo {author} {\bibfnamefont {D.}~\bibnamefont {Meidan}},\ }\bibfield
  {title} {\bibinfo {title} {Localization enhanced and degraded topological
  order in interacting $p$-wave wires},\ }\href
  {https://doi.org/10.1103/PhysRevB.97.085425} {\bibfield  {journal} {\bibinfo
  {journal} {Phys. Rev. B}\ }\textbf {\bibinfo {volume} {97}},\ \bibinfo
  {pages} {085425} (\bibinfo {year} {2018})}\BibitemShut {NoStop}%
\bibitem [{\citenamefont {Agarwala}\ and\ \citenamefont
  {Shenoy}(2017)}]{agarwala2017}%
  \BibitemOpen
  \bibfield  {author} {\bibinfo {author} {\bibfnamefont {A.}~\bibnamefont
  {Agarwala}}\ and\ \bibinfo {author} {\bibfnamefont {V.~B.}\ \bibnamefont
  {Shenoy}},\ }\bibfield  {title} {\bibinfo {title} {Topological insulators in
  amorphous systems},\ }\href {https://doi.org/10.1103/PhysRevLett.118.236402}
  {\bibfield  {journal} {\bibinfo  {journal} {Phys. Rev. Lett.}\ }\textbf
  {\bibinfo {volume} {118}},\ \bibinfo {pages} {236402} (\bibinfo {year}
  {2017})}\BibitemShut {NoStop}%
\bibitem [{\citenamefont {Mansha}\ and\ \citenamefont
  {Chong}(2017)}]{Mansha2017}%
  \BibitemOpen
  \bibfield  {author} {\bibinfo {author} {\bibfnamefont {S.}~\bibnamefont
  {Mansha}}\ and\ \bibinfo {author} {\bibfnamefont {Y.~D.}\ \bibnamefont
  {Chong}},\ }\bibfield  {title} {\bibinfo {title} {Robust edge states in
  amorphous gyromagnetic photonic lattices},\ }\href
  {https://doi.org/10.1103/PhysRevB.96.121405} {\bibfield  {journal} {\bibinfo
  {journal} {Phys. Rev. B}\ }\textbf {\bibinfo {volume} {96}},\ \bibinfo
  {pages} {121405} (\bibinfo {year} {2017})}\BibitemShut {NoStop}%
\bibitem [{\citenamefont {Xiao}\ and\ \citenamefont {Fan}(2017)}]{Xiao2017}%
  \BibitemOpen
  \bibfield  {author} {\bibinfo {author} {\bibfnamefont {M.}~\bibnamefont
  {Xiao}}\ and\ \bibinfo {author} {\bibfnamefont {S.}~\bibnamefont {Fan}},\
  }\bibfield  {title} {\bibinfo {title} {Photonic chern insulator through
  homogenization of an array of particles},\ }\href
  {https://doi.org/10.1103/PhysRevB.96.100202} {\bibfield  {journal} {\bibinfo
  {journal} {Phys. Rev. B}\ }\textbf {\bibinfo {volume} {96}},\ \bibinfo
  {pages} {100202} (\bibinfo {year} {2017})}\BibitemShut {NoStop}%
\bibitem [{\citenamefont {Mitchell}\ \emph {et~al.}(2018)\citenamefont
  {Mitchell}, \citenamefont {Nash}, \citenamefont {Hexner}, \citenamefont
  {Turner},\ and\ \citenamefont {Irvine}}]{mitchell_amorphous_2018}%
  \BibitemOpen
  \bibfield  {author} {\bibinfo {author} {\bibfnamefont {N.~P.}\ \bibnamefont
  {Mitchell}}, \bibinfo {author} {\bibfnamefont {L.~M.}\ \bibnamefont {Nash}},
  \bibinfo {author} {\bibfnamefont {D.}~\bibnamefont {Hexner}}, \bibinfo
  {author} {\bibfnamefont {A.~M.}\ \bibnamefont {Turner}},\ and\ \bibinfo
  {author} {\bibfnamefont {W.~T.~M.}\ \bibnamefont {Irvine}},\ }\bibfield
  {title} {\bibinfo {title} {Amorphous topological insulators constructed from
  random point sets},\ }\href {https://doi.org/10.1038/s41567-017-0024-5}
  {\bibfield  {journal} {\bibinfo  {journal} {Nat. Phys.}\ }\textbf {\bibinfo
  {volume} {14}},\ \bibinfo {pages} {380} (\bibinfo {year} {2018})}\BibitemShut
  {NoStop}%
\bibitem [{\citenamefont {Bourne}\ and\ \citenamefont
  {Prodan}(2018)}]{Bourne:2018jr}%
  \BibitemOpen
  \bibfield  {author} {\bibinfo {author} {\bibfnamefont {C.}~\bibnamefont
  {Bourne}}\ and\ \bibinfo {author} {\bibfnamefont {E.}~\bibnamefont
  {Prodan}},\ }\bibfield  {title} {{\selectlanguage {English}\bibinfo {title}
  {{Non-commutative Chern numbers for generic aperiodic discrete systems}}},\
  }\href {https://doi.org/10.1088/1751-8121/aac093} {\bibfield  {journal}
  {\bibinfo  {journal} {J. Phys. A: Math. Theor.}\ }\textbf {\bibinfo {volume}
  {51}},\ \bibinfo {pages} {235202} (\bibinfo {year} {2018})}\BibitemShut
  {NoStop}%
\bibitem [{\citenamefont {P{\"o}yh{\"o}nen}\ \emph {et~al.}(2018)\citenamefont
  {P{\"o}yh{\"o}nen}, \citenamefont {Sahlberg}, \citenamefont {Weststr{\"o}m},\
  and\ \citenamefont {Ojanen}}]{Poyhonen2018}%
  \BibitemOpen
  \bibfield  {author} {\bibinfo {author} {\bibfnamefont {K.}~\bibnamefont
  {P{\"o}yh{\"o}nen}}, \bibinfo {author} {\bibfnamefont {I.}~\bibnamefont
  {Sahlberg}}, \bibinfo {author} {\bibfnamefont {A.}~\bibnamefont
  {Weststr{\"o}m}},\ and\ \bibinfo {author} {\bibfnamefont {T.}~\bibnamefont
  {Ojanen}},\ }\bibfield  {title} {\bibinfo {title} {Amorphous topological
  superconductivity in a shiba glass},\ }\href
  {https://doi.org/10.1038/s41467-018-04532-x} {\bibfield  {journal} {\bibinfo
  {journal} {Nat. Commun.}\ }\textbf {\bibinfo {volume} {9}},\ \bibinfo {pages}
  {2103} (\bibinfo {year} {2018})}\BibitemShut {NoStop}%
\bibitem [{\citenamefont {Minarelli}\ \emph {et~al.}(2019)\citenamefont
  {Minarelli}, \citenamefont {Pöyhönen}, \citenamefont {van Dalum},
  \citenamefont {Ojanen},\ and\ \citenamefont
  {Fritz}}]{minarelli_engineering_2019}%
  \BibitemOpen
  \bibfield  {author} {\bibinfo {author} {\bibfnamefont {E.~L.}\ \bibnamefont
  {Minarelli}}, \bibinfo {author} {\bibfnamefont {K.}~\bibnamefont
  {Pöyhönen}}, \bibinfo {author} {\bibfnamefont {G.~A.~R.}\ \bibnamefont {van
  Dalum}}, \bibinfo {author} {\bibfnamefont {T.}~\bibnamefont {Ojanen}},\ and\
  \bibinfo {author} {\bibfnamefont {L.}~\bibnamefont {Fritz}},\ }\bibfield
  {title} {\bibinfo {title} {Engineering of {Chern} insulators and circuits of
  topological edge states},\ }\href
  {https://doi.org/10.1103/PhysRevB.99.165413} {\bibfield  {journal} {\bibinfo
  {journal} {Phys. Rev. B}\ }\textbf {\bibinfo {volume} {99}},\ \bibinfo
  {pages} {165413} (\bibinfo {year} {2019})}\BibitemShut {NoStop}%
\bibitem [{\citenamefont {Chern}(2019)}]{chern_topological_2019}%
  \BibitemOpen
  \bibfield  {author} {\bibinfo {author} {\bibfnamefont {G.-W.}\ \bibnamefont
  {Chern}},\ }\bibfield  {title} {\bibinfo {title} {Topological insulator in an
  atomic liquid},\ }\href {https://doi.org/10.1209/0295-5075/126/37002}
  {\bibfield  {journal} {\bibinfo  {journal} {Europhys. Lett.}\ }\textbf
  {\bibinfo {volume} {126}},\ \bibinfo {pages} {37002} (\bibinfo {year}
  {2019})}\BibitemShut {NoStop}%
\bibitem [{\citenamefont {Mano}\ and\ \citenamefont
  {Ohtsuki}(2019)}]{mano_application_2019}%
  \BibitemOpen
  \bibfield  {author} {\bibinfo {author} {\bibfnamefont {T.}~\bibnamefont
  {Mano}}\ and\ \bibinfo {author} {\bibfnamefont {T.}~\bibnamefont {Ohtsuki}},\
  }\bibfield  {title} {\bibinfo {title} {Application of {Convolutional}
  {Neural} {Network} to {Quantum} {Percolation} in {Topological}
  {Insulators}},\ }\href {https://doi.org/10.7566/JPSJ.88.123704} {\bibfield
  {journal} {\bibinfo  {journal} {JPSJ}\ }\textbf {\bibinfo {volume} {88}},\
  \bibinfo {pages} {123704} (\bibinfo {year} {2019})}\BibitemShut {NoStop}%
\bibitem [{\citenamefont {Costa}\ \emph {et~al.}(2019)\citenamefont {Costa},
  \citenamefont {Schleder}, \citenamefont {Buongiorno~Nardelli}, \citenamefont
  {Lewenkopf},\ and\ \citenamefont {Fazzio}}]{Costa:2019kc}%
  \BibitemOpen
  \bibfield  {author} {\bibinfo {author} {\bibfnamefont {M.}~\bibnamefont
  {Costa}}, \bibinfo {author} {\bibfnamefont {G.~R.}\ \bibnamefont {Schleder}},
  \bibinfo {author} {\bibfnamefont {M.}~\bibnamefont {Buongiorno~Nardelli}},
  \bibinfo {author} {\bibfnamefont {C.}~\bibnamefont {Lewenkopf}},\ and\
  \bibinfo {author} {\bibfnamefont {A.}~\bibnamefont {Fazzio}},\ }\bibfield
  {title} {\bibinfo {title} {Toward realistic amorphous topological
  insulators},\ }\href {https://doi.org/10.1021/acs.nanolett.9b03881}
  {\bibfield  {journal} {\bibinfo  {journal} {Nano Lett.}\ }\textbf {\bibinfo
  {volume} {19}},\ \bibinfo {pages} {8941} (\bibinfo {year}
  {2019})}\BibitemShut {NoStop}%
\bibitem [{\citenamefont {Marsal}\ \emph {et~al.}(2020)\citenamefont {Marsal},
  \citenamefont {Varjas},\ and\ \citenamefont
  {Grushin}}]{marsal_topological_2020}%
  \BibitemOpen
  \bibfield  {author} {\bibinfo {author} {\bibfnamefont {Q.}~\bibnamefont
  {Marsal}}, \bibinfo {author} {\bibfnamefont {D.}~\bibnamefont {Varjas}},\
  and\ \bibinfo {author} {\bibfnamefont {A.~G.}\ \bibnamefont {Grushin}},\
  }\bibfield  {title} {\bibinfo {title} {Topological {Weaire}–{Thorpe} models
  of amorphous matter},\ }\href {https://doi.org/10.1073/pnas.2007384117}
  {\bibfield  {journal} {\bibinfo  {journal} {PNAS}\ }\textbf {\bibinfo
  {volume} {117}},\ \bibinfo {pages} {30260} (\bibinfo {year}
  {2020})}\BibitemShut {NoStop}%
\bibitem [{\citenamefont {Sahlberg}\ \emph {et~al.}(2020)\citenamefont
  {Sahlberg}, \citenamefont {Weststr\"om}, \citenamefont {P\"oyh\"onen},\ and\
  \citenamefont {Ojanen}}]{Sahlberg2020}%
  \BibitemOpen
  \bibfield  {author} {\bibinfo {author} {\bibfnamefont {I.}~\bibnamefont
  {Sahlberg}}, \bibinfo {author} {\bibfnamefont {A.}~\bibnamefont
  {Weststr\"om}}, \bibinfo {author} {\bibfnamefont {K.}~\bibnamefont
  {P\"oyh\"onen}},\ and\ \bibinfo {author} {\bibfnamefont {T.}~\bibnamefont
  {Ojanen}},\ }\bibfield  {title} {\bibinfo {title} {Topological phase
  transitions in glassy quantum matter},\ }\href
  {https://doi.org/10.1103/PhysRevResearch.2.013053} {\bibfield  {journal}
  {\bibinfo  {journal} {Phys. Rev. Res.}\ }\textbf {\bibinfo {volume} {2}},\
  \bibinfo {pages} {013053} (\bibinfo {year} {2020})}\BibitemShut {NoStop}%
\bibitem [{\citenamefont {Ivaki}\ \emph {et~al.}(2020)\citenamefont {Ivaki},
  \citenamefont {Sahlberg},\ and\ \citenamefont
  {Ojanen}}]{ivaki_criticality_2020}%
  \BibitemOpen
  \bibfield  {author} {\bibinfo {author} {\bibfnamefont {M.~N.}\ \bibnamefont
  {Ivaki}}, \bibinfo {author} {\bibfnamefont {I.}~\bibnamefont {Sahlberg}},\
  and\ \bibinfo {author} {\bibfnamefont {T.}~\bibnamefont {Ojanen}},\
  }\bibfield  {title} {\bibinfo {title} {Criticality in amorphous topological
  matter: {Beyond} the universal scaling paradigm},\ }\href
  {https://doi.org/10.1103/PhysRevResearch.2.043301} {\bibfield  {journal}
  {\bibinfo  {journal} {Phys. Rev. Res.}\ }\textbf {\bibinfo {volume} {2}},\
  \bibinfo {pages} {043301} (\bibinfo {year} {2020})}\BibitemShut {NoStop}%
\bibitem [{\citenamefont {Wang}\ \emph {et~al.}(2021)\citenamefont {Wang},
  \citenamefont {Yang}, \citenamefont {Dai},\ and\ \citenamefont
  {Xu}}]{wang_structural-disorder-induced_2021}%
  \BibitemOpen
  \bibfield  {author} {\bibinfo {author} {\bibfnamefont {J.-H.}\ \bibnamefont
  {Wang}}, \bibinfo {author} {\bibfnamefont {Y.-B.}\ \bibnamefont {Yang}},
  \bibinfo {author} {\bibfnamefont {N.}~\bibnamefont {Dai}},\ and\ \bibinfo
  {author} {\bibfnamefont {Y.}~\bibnamefont {Xu}},\ }\bibfield  {title}
  {\bibinfo {title} {Structural-{Disorder}-{Induced} {Second}-{Order}
  {Topological} {Insulators} in {Three} {Dimensions}},\ }\href
  {https://doi.org/10.1103/PhysRevLett.126.206404} {\bibfield  {journal}
  {\bibinfo  {journal} {Phys. Rev. Lett.}\ }\textbf {\bibinfo {volume} {126}},\
  \bibinfo {pages} {206404} (\bibinfo {year} {2021})}\BibitemShut {NoStop}%
\bibitem [{\citenamefont {Focassio}\ \emph {et~al.}(2021)\citenamefont
  {Focassio}, \citenamefont {Schleder}, \citenamefont {Costa}, \citenamefont
  {Fazzio},\ and\ \citenamefont {Lewenkopf}}]{focassio_structural_2021}%
  \BibitemOpen
  \bibfield  {author} {\bibinfo {author} {\bibfnamefont {B.}~\bibnamefont
  {Focassio}}, \bibinfo {author} {\bibfnamefont {G.~R.}\ \bibnamefont
  {Schleder}}, \bibinfo {author} {\bibfnamefont {M.}~\bibnamefont {Costa}},
  \bibinfo {author} {\bibfnamefont {A.}~\bibnamefont {Fazzio}},\ and\ \bibinfo
  {author} {\bibfnamefont {C.}~\bibnamefont {Lewenkopf}},\ }\bibfield  {title}
  {\bibinfo {title} {Structural and electronic properties of realistic
  two-dimensional amorphous topological insulators},\ }\href
  {https://doi.org/10.1088/2053-1583/abdb97} {\bibfield  {journal} {\bibinfo
  {journal} {2d Mater.}\ }\textbf {\bibinfo {volume} {8}},\ \bibinfo {pages}
  {025032} (\bibinfo {year} {2021})}\BibitemShut {NoStop}%
\bibitem [{\citenamefont {Mitchell}\ \emph {et~al.}(2021)\citenamefont
  {Mitchell}, \citenamefont {Turner},\ and\ \citenamefont
  {Irvine}}]{Mitchell2021}%
  \BibitemOpen
  \bibfield  {author} {\bibinfo {author} {\bibfnamefont {N.~P.}\ \bibnamefont
  {Mitchell}}, \bibinfo {author} {\bibfnamefont {A.~M.}\ \bibnamefont
  {Turner}},\ and\ \bibinfo {author} {\bibfnamefont {W.~T.~M.}\ \bibnamefont
  {Irvine}},\ }\bibfield  {title} {\bibinfo {title} {Real-space origin of
  topological band gaps, localization, and reentrant phase transitions in
  gyroscopic metamaterials},\ }\href
  {https://doi.org/10.1103/PhysRevE.104.025007} {\bibfield  {journal} {\bibinfo
   {journal} {Phys. Rev. E}\ }\textbf {\bibinfo {volume} {104}},\ \bibinfo
  {pages} {025007} (\bibinfo {year} {2021})}\BibitemShut {NoStop}%
\bibitem [{\citenamefont {Spring}\ \emph {et~al.}(2021)\citenamefont {Spring},
  \citenamefont {Akhmerov},\ and\ \citenamefont
  {Varjas}}]{spring_amorphous_2021}%
  \BibitemOpen
  \bibfield  {author} {\bibinfo {author} {\bibfnamefont {H.}~\bibnamefont
  {Spring}}, \bibinfo {author} {\bibfnamefont {A.}~\bibnamefont {Akhmerov}},\
  and\ \bibinfo {author} {\bibfnamefont {D.}~\bibnamefont {Varjas}},\
  }\bibfield  {title} {\bibinfo {title} {Amorphous topological phases protected
  by continuous rotation symmetry},\ }\href
  {https://doi.org/10.21468/SciPostPhys.11.2.022} {\bibfield  {journal}
  {\bibinfo  {journal} {SciPost Phys.}\ }\textbf {\bibinfo {volume} {11}},\
  \bibinfo {pages} {022} (\bibinfo {year} {2021})}\BibitemShut {NoStop}%
\bibitem [{\citenamefont {Wang}\ \emph {et~al.}(2022)\citenamefont {Wang},
  \citenamefont {Cheng}, \citenamefont {Liu}, \citenamefont {Liu},\ and\
  \citenamefont {Huang}}]{wang_structural_2022}%
  \BibitemOpen
  \bibfield  {author} {\bibinfo {author} {\bibfnamefont {C.}~\bibnamefont
  {Wang}}, \bibinfo {author} {\bibfnamefont {T.}~\bibnamefont {Cheng}},
  \bibinfo {author} {\bibfnamefont {Z.}~\bibnamefont {Liu}}, \bibinfo {author}
  {\bibfnamefont {F.}~\bibnamefont {Liu}},\ and\ \bibinfo {author}
  {\bibfnamefont {H.}~\bibnamefont {Huang}},\ }\bibfield  {title} {\bibinfo
  {title} {Structural {Amorphization}-{Induced} {Topological} {Order}},\ }\href
  {https://doi.org/10.1103/PhysRevLett.128.056401} {\bibfield  {journal}
  {\bibinfo  {journal} {Phys. Rev. Lett.}\ }\textbf {\bibinfo {volume} {128}},\
  \bibinfo {pages} {056401} (\bibinfo {year} {2022})}\BibitemShut {NoStop}%
\bibitem [{\citenamefont {Ur\'{\i}a-\'Alvarez}\ \emph
  {et~al.}(2022)\citenamefont {Ur\'{\i}a-\'Alvarez}, \citenamefont
  {Molpeceres-Mingo},\ and\ \citenamefont {Palacios}}]{uria-alvarez_deep_2022}%
  \BibitemOpen
  \bibfield  {author} {\bibinfo {author} {\bibfnamefont {A.~J.}\ \bibnamefont
  {Ur\'{\i}a-\'Alvarez}}, \bibinfo {author} {\bibfnamefont {D.}~\bibnamefont
  {Molpeceres-Mingo}},\ and\ \bibinfo {author} {\bibfnamefont {J.~J.}\
  \bibnamefont {Palacios}},\ }\bibfield  {title} {\bibinfo {title} {Deep
  learning for disordered topological insulators through their entanglement
  spectrum},\ }\href {https://doi.org/10.1103/PhysRevB.105.155128} {\bibfield
  {journal} {\bibinfo  {journal} {Phys. Rev. B}\ }\textbf {\bibinfo {volume}
  {105}},\ \bibinfo {pages} {155128} (\bibinfo {year} {2022})}\BibitemShut
  {NoStop}%
\bibitem [{\citenamefont {Mu\~noz Segovia}\ \emph {et~al.}(2023)\citenamefont
  {Mu\~noz Segovia}, \citenamefont {Corbae}, \citenamefont {Varjas},
  \citenamefont {Hellman}, \citenamefont {Griffin},\ and\ \citenamefont
  {Grushin}}]{spillage_2022}%
  \BibitemOpen
  \bibfield  {author} {\bibinfo {author} {\bibfnamefont {D.}~\bibnamefont
  {Mu\~noz Segovia}}, \bibinfo {author} {\bibfnamefont {P.}~\bibnamefont
  {Corbae}}, \bibinfo {author} {\bibfnamefont {D.}~\bibnamefont {Varjas}},
  \bibinfo {author} {\bibfnamefont {F.}~\bibnamefont {Hellman}}, \bibinfo
  {author} {\bibfnamefont {S.~M.}\ \bibnamefont {Griffin}},\ and\ \bibinfo
  {author} {\bibfnamefont {A.~G.}\ \bibnamefont {Grushin}},\ }\bibfield
  {title} {\bibinfo {title} {Structural spillage: An efficient method to
  identify noncrystalline topological materials},\ }\href
  {https://doi.org/10.1103/PhysRevResearch.5.L042011} {\bibfield  {journal}
  {\bibinfo  {journal} {Phys. Rev. Res.}\ }\textbf {\bibinfo {volume} {5}},\
  \bibinfo {pages} {L042011} (\bibinfo {year} {2023})}\BibitemShut {NoStop}%
\bibitem [{\citenamefont {Cassella}\ \emph {et~al.}(2023)\citenamefont
  {Cassella}, \citenamefont {d'Ornellas}, \citenamefont {Hodson}, \citenamefont
  {Natori},\ and\ \citenamefont {Knolle}}]{Cassella2023}%
  \BibitemOpen
  \bibfield  {author} {\bibinfo {author} {\bibfnamefont {G.}~\bibnamefont
  {Cassella}}, \bibinfo {author} {\bibfnamefont {P.}~\bibnamefont
  {d'Ornellas}}, \bibinfo {author} {\bibfnamefont {T.}~\bibnamefont {Hodson}},
  \bibinfo {author} {\bibfnamefont {W.~M.~H.}\ \bibnamefont {Natori}},\ and\
  \bibinfo {author} {\bibfnamefont {J.}~\bibnamefont {Knolle}},\ }\bibfield
  {title} {\bibinfo {title} {An exact chiral amorphous spin liquid},\ }\href
  {https://doi.org/10.1038/s41467-023-42105-9} {\bibfield  {journal} {\bibinfo
  {journal} {Nat. Commun.}\ }\textbf {\bibinfo {volume} {14}},\ \bibinfo
  {pages} {6663} (\bibinfo {year} {2023})}\BibitemShut {NoStop}%
\bibitem [{\citenamefont {Grushin}\ and\ \citenamefont
  {Repellin}(2023)}]{Grushin2023}%
  \BibitemOpen
  \bibfield  {author} {\bibinfo {author} {\bibfnamefont {A.~G.}\ \bibnamefont
  {Grushin}}\ and\ \bibinfo {author} {\bibfnamefont {C.}~\bibnamefont
  {Repellin}},\ }\bibfield  {title} {\bibinfo {title} {Amorphous and
  polycrystalline routes toward a chiral spin liquid},\ }\href
  {https://doi.org/10.1103/PhysRevLett.130.186702} {\bibfield  {journal}
  {\bibinfo  {journal} {Phys. Rev. Lett.}\ }\textbf {\bibinfo {volume} {130}},\
  \bibinfo {pages} {186702} (\bibinfo {year} {2023})}\BibitemShut {NoStop}%
\bibitem [{\citenamefont {Corbae}\ \emph {et~al.}(2023)\citenamefont {Corbae},
  \citenamefont {Hannukainen}, \citenamefont {Marsal}, \citenamefont
  {Muñoz-Segovia},\ and\ \citenamefont {Grushin}}]{Corbae2023}%
  \BibitemOpen
  \bibfield  {author} {\bibinfo {author} {\bibfnamefont {P.}~\bibnamefont
  {Corbae}}, \bibinfo {author} {\bibfnamefont {J.~D.}\ \bibnamefont
  {Hannukainen}}, \bibinfo {author} {\bibfnamefont {Q.}~\bibnamefont {Marsal}},
  \bibinfo {author} {\bibfnamefont {D.}~\bibnamefont {Muñoz-Segovia}},\ and\
  \bibinfo {author} {\bibfnamefont {A.~G.}\ \bibnamefont {Grushin}},\
  }\bibfield  {title} {\bibinfo {title} {Amorphous topological matter: Theory
  and experiment},\ }\href {https://doi.org/10.1209/0295-5075/acc2e2}
  {\bibfield  {journal} {\bibinfo  {journal} {EPL}\ }\textbf {\bibinfo {volume}
  {142}},\ \bibinfo {pages} {16001} (\bibinfo {year} {2023})}\BibitemShut
  {NoStop}%
\bibitem [{\citenamefont {Manna}\ \emph {et~al.}(2024)\citenamefont {Manna},
  \citenamefont {Das},\ and\ \citenamefont {Roy}}]{Manna2024}%
  \BibitemOpen
  \bibfield  {author} {\bibinfo {author} {\bibfnamefont {S.}~\bibnamefont
  {Manna}}, \bibinfo {author} {\bibfnamefont {S.~K.}\ \bibnamefont {Das}},\
  and\ \bibinfo {author} {\bibfnamefont {B.}~\bibnamefont {Roy}},\ }\bibfield
  {title} {\bibinfo {title} {Noncrystalline topological superconductors},\
  }\href {https://doi.org/10.1103/PhysRevB.109.174512} {\bibfield  {journal}
  {\bibinfo  {journal} {Phys. Rev. B}\ }\textbf {\bibinfo {volume} {109}},\
  \bibinfo {pages} {174512} (\bibinfo {year} {2024})}\BibitemShut {NoStop}%
\bibitem [{\citenamefont {Marsal}\ \emph {et~al.}(2023)\citenamefont {Marsal},
  \citenamefont {Varjas},\ and\ \citenamefont
  {Grushin}}]{marsal_obstructed_2022}%
  \BibitemOpen
  \bibfield  {author} {\bibinfo {author} {\bibfnamefont {Q.}~\bibnamefont
  {Marsal}}, \bibinfo {author} {\bibfnamefont {D.}~\bibnamefont {Varjas}},\
  and\ \bibinfo {author} {\bibfnamefont {A.~G.}\ \bibnamefont {Grushin}},\
  }\bibfield  {title} {\bibinfo {title} {Obstructed insulators and flat bands
  in topological phase-change materials},\ }\href
  {https://doi.org/10.1103/PhysRevB.107.045119} {\bibfield  {journal} {\bibinfo
   {journal} {Phys. Rev. B}\ }\textbf {\bibinfo {volume} {107}},\ \bibinfo
  {pages} {045119} (\bibinfo {year} {2023})}\BibitemShut {NoStop}%
\bibitem [{\citenamefont {Ur\'{\i}a-\'Alvarez}\ and\ \citenamefont
  {Palacios}(2025)}]{uria2024amorphization}%
  \BibitemOpen
  \bibfield  {author} {\bibinfo {author} {\bibfnamefont {A.~J.}\ \bibnamefont
  {Ur\'{\i}a-\'Alvarez}}\ and\ \bibinfo {author} {\bibfnamefont {J.~J.}\
  \bibnamefont {Palacios}},\ }\bibfield  {title} {\bibinfo {title}
  {Amorphization-induced topological and insulator-metal transitions in
  bidimensional ${\mathrm{bi}}_{x}{\mathrm{sb}}_{1\ensuremath{-}x}$ alloys},\
  }\href {https://doi.org/10.1103/z5tm-cvgn} {\bibfield  {journal} {\bibinfo
  {journal} {Phys. Rev. Res.}\ }\textbf {\bibinfo {volume} {7}},\ \bibinfo
  {pages} {043263} (\bibinfo {year} {2025})}\BibitemShut {NoStop}%
\bibitem [{\citenamefont {Alexandradinata}\ \emph {et~al.}(2014)\citenamefont
  {Alexandradinata}, \citenamefont {Dai},\ and\ \citenamefont
  {Bernevig}}]{alexandradinata2014Wilson}%
  \BibitemOpen
  \bibfield  {author} {\bibinfo {author} {\bibfnamefont {A.}~\bibnamefont
  {Alexandradinata}}, \bibinfo {author} {\bibfnamefont {X.}~\bibnamefont
  {Dai}},\ and\ \bibinfo {author} {\bibfnamefont {B.~A.}\ \bibnamefont
  {Bernevig}},\ }\bibfield  {title} {\bibinfo {title} {Wilson-loop
  characterization of inversion-symmetric topological insulators},\ }\href
  {https://doi.org/10.1103/PhysRevB.89.155114} {\bibfield  {journal} {\bibinfo
  {journal} {Phys. Rev. B}\ }\textbf {\bibinfo {volume} {89}},\ \bibinfo
  {pages} {155114} (\bibinfo {year} {2014})}\BibitemShut {NoStop}%
\bibitem [{\citenamefont {Benalcazar}\ \emph
  {et~al.}(2017{\natexlab{b}})\citenamefont {Benalcazar}, \citenamefont
  {Bernevig},\ and\ \citenamefont {Hughes}}]{Benalcazar2017electric}%
  \BibitemOpen
  \bibfield  {author} {\bibinfo {author} {\bibfnamefont {W.~A.}\ \bibnamefont
  {Benalcazar}}, \bibinfo {author} {\bibfnamefont {B.~A.}\ \bibnamefont
  {Bernevig}},\ and\ \bibinfo {author} {\bibfnamefont {T.~L.}\ \bibnamefont
  {Hughes}},\ }\bibfield  {title} {\bibinfo {title} {Electric multipole
  moments, topological multipole moment pumping, and chiral hinge states in
  crystalline insulators},\ }\href {https://doi.org/10.1103/PhysRevB.96.245115}
  {\bibfield  {journal} {\bibinfo  {journal} {Phys. Rev. B}\ }\textbf {\bibinfo
  {volume} {96}},\ \bibinfo {pages} {245115} (\bibinfo {year}
  {2017}{\natexlab{b}})}\BibitemShut {NoStop}%
\bibitem [{\citenamefont {Zijderveld}\ \emph {et~al.}(2025)\citenamefont
  {Zijderveld}, \citenamefont {Day},\ and\ \citenamefont
  {Akhmerov}}]{zijderveld2025scatteringtheoryhigherorder}%
  \BibitemOpen
  \bibfield  {author} {\bibinfo {author} {\bibfnamefont {R.~J.}\ \bibnamefont
  {Zijderveld}}, \bibinfo {author} {\bibfnamefont {I.~A.}\ \bibnamefont
  {Day}},\ and\ \bibinfo {author} {\bibfnamefont {A.~R.}\ \bibnamefont
  {Akhmerov}},\ }\bibfield  {title} {\bibinfo {title} {{Scattering theory of
  higher order topological phases}},\ }\href
  {https://doi.org/10.21468/SciPostPhys.19.2.058} {\bibfield  {journal}
  {\bibinfo  {journal} {SciPost Phys.}\ }\textbf {\bibinfo {volume} {19}},\
  \bibinfo {pages} {058} (\bibinfo {year} {2025})}\BibitemShut {NoStop}%
\bibitem [{\citenamefont {Cerjan}\ \emph {et~al.}(2024)\citenamefont {Cerjan},
  \citenamefont {Loring},\ and\ \citenamefont {Schulz-Baldes}}]{CerjanHOTI}%
  \BibitemOpen
  \bibfield  {author} {\bibinfo {author} {\bibfnamefont {A.}~\bibnamefont
  {Cerjan}}, \bibinfo {author} {\bibfnamefont {T.~A.}\ \bibnamefont {Loring}},\
  and\ \bibinfo {author} {\bibfnamefont {H.}~\bibnamefont {Schulz-Baldes}},\
  }\bibfield  {title} {\bibinfo {title} {Local markers for crystalline
  topology},\ }\href {https://doi.org/10.1103/PhysRevLett.132.073803}
  {\bibfield  {journal} {\bibinfo  {journal} {Phys. Rev. Lett.}\ }\textbf
  {\bibinfo {volume} {132}},\ \bibinfo {pages} {073803} (\bibinfo {year}
  {2024})}\BibitemShut {NoStop}%
\bibitem [{\citenamefont {Jezequel}\ and\ \citenamefont
  {Delplace}(2024)}]{Jezequel2024}%
  \BibitemOpen
  \bibfield  {author} {\bibinfo {author} {\bibfnamefont {L.}~\bibnamefont
  {Jezequel}}\ and\ \bibinfo {author} {\bibfnamefont {P.}~\bibnamefont
  {Delplace}},\ }\bibfield  {title} {\bibinfo {title} {{Mode-shell
  correspondence, a unifying phase space theory in topological physics - Part
  I: Chiral number of zero-modes}},\ }\href
  {https://doi.org/10.21468/SciPostPhys.17.2.060} {\bibfield  {journal}
  {\bibinfo  {journal} {SciPost Phys.}\ }\textbf {\bibinfo {volume} {17}},\
  \bibinfo {pages} {060} (\bibinfo {year} {2024})}\BibitemShut {NoStop}%
\bibitem [{\citenamefont {Jezequel}\ and\ \citenamefont
  {Delplace}(2025)}]{Jezequel2025}%
  \BibitemOpen
  \bibfield  {author} {\bibinfo {author} {\bibfnamefont {L.}~\bibnamefont
  {Jezequel}}\ and\ \bibinfo {author} {\bibfnamefont {P.}~\bibnamefont
  {Delplace}},\ }\bibfield  {title} {\bibinfo {title} {{Mode-Shell
  correspondence, a unifying phase space theory in topological physics - Part
  II: Higher-dimensional spectral invariants}},\ }\href
  {https://doi.org/10.21468/SciPostPhys.18.6.193} {\bibfield  {journal}
  {\bibinfo  {journal} {SciPost Phys.}\ }\textbf {\bibinfo {volume} {18}},\
  \bibinfo {pages} {193} (\bibinfo {year} {2025})}\BibitemShut {NoStop}%
\bibitem [{\citenamefont {Kane}\ and\ \citenamefont
  {Lubensky}(2014)}]{kaneLubenski}%
  \BibitemOpen
  \bibfield  {author} {\bibinfo {author} {\bibfnamefont {C.~L.}\ \bibnamefont
  {Kane}}\ and\ \bibinfo {author} {\bibfnamefont {T.~C.}\ \bibnamefont
  {Lubensky}},\ }\bibfield  {title} {\bibinfo {title} {Topological boundary
  modes in isostatic lattices},\ }\href {https://doi.org/10.1038/nphys2835}
  {\bibfield  {journal} {\bibinfo  {journal} {Nat. Phys.}\ }\textbf {\bibinfo
  {volume} {10}},\ \bibinfo {pages} {39} (\bibinfo {year} {2014})}\BibitemShut
  {NoStop}%
\bibitem [{\citenamefont {Guzmán}\ \emph {et~al.}(2022)\citenamefont
  {Guzmán}, \citenamefont {Bartolo},\ and\ \citenamefont
  {Carpentier}}]{MarceloTango}%
  \BibitemOpen
  \bibfield  {author} {\bibinfo {author} {\bibfnamefont {M.}~\bibnamefont
  {Guzmán}}, \bibinfo {author} {\bibfnamefont {D.}~\bibnamefont {Bartolo}},\
  and\ \bibinfo {author} {\bibfnamefont {D.}~\bibnamefont {Carpentier}},\
  }\bibfield  {title} {\bibinfo {title} {{Geometry and topology tango in
  ordered and amorphous chiral matter}},\ }\href
  {https://doi.org/10.21468/SciPostPhys.12.1.038} {\bibfield  {journal}
  {\bibinfo  {journal} {SciPost Phys.}\ }\textbf {\bibinfo {volume} {12}},\
  \bibinfo {pages} {038} (\bibinfo {year} {2022})}\BibitemShut {NoStop}%
\bibitem [{\citenamefont {Yang}\ \emph {et~al.}(2022)\citenamefont {Yang},
  \citenamefont {Song}, \citenamefont {Cao},\ and\ \citenamefont
  {Yan}}]{yang_experimental_2022}%
  \BibitemOpen
  \bibfield  {author} {\bibinfo {author} {\bibfnamefont {H.}~\bibnamefont
  {Yang}}, \bibinfo {author} {\bibfnamefont {L.}~\bibnamefont {Song}}, \bibinfo
  {author} {\bibfnamefont {Y.}~\bibnamefont {Cao}},\ and\ \bibinfo {author}
  {\bibfnamefont {P.}~\bibnamefont {Yan}},\ }\bibfield  {title} {\bibinfo
  {title} {Experimental {Realization} of {Two}-{Dimensional} {Weak}
  {Topological} {Insulators}},\ }\href
  {https://doi.org/10.1021/acs.nanolett.2c00555} {\bibfield  {journal}
  {\bibinfo  {journal} {Nano Lett.}\ }\textbf {\bibinfo {volume} {22}},\
  \bibinfo {pages} {3125} (\bibinfo {year} {2022})},\ \bibinfo {note}
  {publisher: American Chemical Society}\BibitemShut {NoStop}%
\bibitem [{\citenamefont {Ringel}\ \emph {et~al.}(2012)\citenamefont {Ringel},
  \citenamefont {Kraus},\ and\ \citenamefont {Stern}}]{StrongsideWeakTI}%
  \BibitemOpen
  \bibfield  {author} {\bibinfo {author} {\bibfnamefont {Z.}~\bibnamefont
  {Ringel}}, \bibinfo {author} {\bibfnamefont {Y.~E.}\ \bibnamefont {Kraus}},\
  and\ \bibinfo {author} {\bibfnamefont {A.}~\bibnamefont {Stern}},\ }\bibfield
   {title} {\bibinfo {title} {Strong side of weak topological insulators},\
  }\href {https://doi.org/10.1103/PhysRevB.86.045102} {\bibfield  {journal}
  {\bibinfo  {journal} {Phys. Rev. B}\ }\textbf {\bibinfo {volume} {86}},\
  \bibinfo {pages} {045102} (\bibinfo {year} {2012})}\BibitemShut {NoStop}%
\bibitem [{\citenamefont {Kobayashi}\ \emph {et~al.}(2013)\citenamefont
  {Kobayashi}, \citenamefont {Ohtsuki},\ and\ \citenamefont
  {Imura}}]{Disorweakandstrong}%
  \BibitemOpen
  \bibfield  {author} {\bibinfo {author} {\bibfnamefont {K.}~\bibnamefont
  {Kobayashi}}, \bibinfo {author} {\bibfnamefont {T.}~\bibnamefont {Ohtsuki}},\
  and\ \bibinfo {author} {\bibfnamefont {K.-I.}\ \bibnamefont {Imura}},\
  }\bibfield  {title} {\bibinfo {title} {Disordered weak and strong topological
  insulators},\ }\href {https://doi.org/10.1103/PhysRevLett.110.236803}
  {\bibfield  {journal} {\bibinfo  {journal} {Phys. Rev. Lett.}\ }\textbf
  {\bibinfo {volume} {110}},\ \bibinfo {pages} {236803} (\bibinfo {year}
  {2013})}\BibitemShut {NoStop}%
\bibitem [{\citenamefont {Fulga}\ \emph {et~al.}(2016)\citenamefont {Fulga},
  \citenamefont {Pikulin},\ and\ \citenamefont {Loring}}]{AperiodicWTI}%
  \BibitemOpen
  \bibfield  {author} {\bibinfo {author} {\bibfnamefont {I.~C.}\ \bibnamefont
  {Fulga}}, \bibinfo {author} {\bibfnamefont {D.~I.}\ \bibnamefont {Pikulin}},\
  and\ \bibinfo {author} {\bibfnamefont {T.~A.}\ \bibnamefont {Loring}},\
  }\bibfield  {title} {\bibinfo {title} {Aperiodic weak topological
  superconductors},\ }\href {https://doi.org/10.1103/PhysRevLett.116.257002}
  {\bibfield  {journal} {\bibinfo  {journal} {Phys. Rev. Lett.}\ }\textbf
  {\bibinfo {volume} {116}},\ \bibinfo {pages} {257002} (\bibinfo {year}
  {2016})}\BibitemShut {NoStop}%
\bibitem [{\citenamefont {Yan}\ \emph {et~al.}(2012)\citenamefont {Yan},
  \citenamefont {M\"uchler},\ and\ \citenamefont {Felser}}]{PredictionWeak}%
  \BibitemOpen
  \bibfield  {author} {\bibinfo {author} {\bibfnamefont {B.}~\bibnamefont
  {Yan}}, \bibinfo {author} {\bibfnamefont {L.}~\bibnamefont {M\"uchler}},\
  and\ \bibinfo {author} {\bibfnamefont {C.}~\bibnamefont {Felser}},\
  }\bibfield  {title} {\bibinfo {title} {Prediction of weak topological
  insulators in layered semiconductors},\ }\href
  {https://doi.org/10.1103/PhysRevLett.109.116406} {\bibfield  {journal}
  {\bibinfo  {journal} {Phys. Rev. Lett.}\ }\textbf {\bibinfo {volume} {109}},\
  \bibinfo {pages} {116406} (\bibinfo {year} {2012})}\BibitemShut {NoStop}%
\bibitem [{\citenamefont {Martínez}\ \emph {et~al.}(2025)\citenamefont
  {Martínez}, \citenamefont {Jezequel}, \citenamefont {Bardarson},
  \citenamefont {Kvorning},\ and\ \citenamefont {Hannukainen}}]{zenodo-repo}%
  \BibitemOpen
  \bibfield  {author} {\bibinfo {author} {\bibfnamefont {M.~F.}\ \bibnamefont
  {Martínez}}, \bibinfo {author} {\bibfnamefont {L.}~\bibnamefont {Jezequel}},
  \bibinfo {author} {\bibfnamefont {J.~H.}\ \bibnamefont {Bardarson}}, \bibinfo
  {author} {\bibfnamefont {T.~K.}\ \bibnamefont {Kvorning}},\ and\ \bibinfo
  {author} {\bibfnamefont {J.~D.}\ \bibnamefont {Hannukainen}},\ }\href
  {https://doi.org/10.5281/zenodo.17175143} {\bibinfo {title} {A one-particle
  density matrix framework for mode-shell correspondence: Characterizing
  topology in amorphous higher-order topological insulators}} (\bibinfo {year}
  {2025})\BibitemShut {NoStop}%
\end{thebibliography}%

\end{document}